\documentclass[a4paper,11pt]{article}
\pdfoutput=1 

\usepackage{jheppub} 

\usepackage[T1]{fontenc} 
\usepackage{graphicx}
\usepackage{amsmath, amsfonts}

\usepackage{amssymb}  
\usepackage{graphicx,subfigure}
\usepackage{exscale}
\usepackage{textcomp}
\usepackage{enumerate}
\usepackage [english]{babel}
\usepackage [autostyle, english = american]{csquotes}
\MakeOuterQuote{"}

\usepackage{color}

\usepackage[normalem]{ulem}  

\newcommand{\non}{\nonumber\\}

\newcommand{\be}{\begin{equation}}
\newcommand{\ee}{\end{equation}}
\newcommand{\bea}{\begin{eqnarray}}
\newcommand{\eea}{\end{eqnarray}}
\newcommand{\ba}[1]{\begin{array}{#1}}
\newcommand{\ea}{\end{array}}

\newcommand{\Tr}{{\rm Tr}}

\title{Towards a holographic quark-hadron continuity}

\author[a]{Kazem Bitaghsir Fadafan,}\author[a]{Farideh Kazemian,}\author[b]{and Andreas Schmitt}
 
\affiliation[a]{Faculty of Physics, Shahrood University of Technology, P.O.Box 3619995161 Shahrood, Iran} 
\affiliation[b]{Mathematical Sciences and STAG Research Centre, University of Southampton, Southampton SO17 1BJ, United Kingdom}

\emailAdd{bitaghsir@shahroodut.ac.ir}
\emailAdd{fkazemian@shahroodut.ac.ir}
\emailAdd{a.schmitt@soton.ac.uk}

\date{21 November 2018}

\abstract{We study dense nuclear and quark matter within a single microscopic approach, namely the 
holographic Sakai-Sugimoto model. Nuclear matter is described via instantons in the bulk, and we show that 
instanton interactions are crucial for a continuous connection of chirally broken and chirally symmetric phases. 
The continuous path from nuclear to quark matter includes metastable and unstable stationary points of the 
potential, while the actual chiral phase transition remains of first order, as in earlier approximations. 
We show that the model parameters can be chosen to reproduce low-density properties of nuclear matter and observe a non-monotonic behavior of the speed of sound as a function of the baryon chemical potential, as suggested by constraints from QCD and astrophysics. 
}

\begin{document}
\maketitle
\flushbottom

\section{Introduction}

\subsection{Context and purpose}

If nuclear matter is compressed to sufficiently large densities it turns into quark matter, which is weakly interacting at asymptotically large densities. The nature of this transition is unknown because first-principle calculations
within Quantum Chromodynamics (QCD) are currently inaccessible for matter at low temperature and large baryon density
due to the strong-coupling nature of the problem and the so-called sign problem of lattice gauge theory \cite{Aarts:2015tyj}. In a simple picture, extreme compression makes the nucleons overlap, and at some point it is the quarks, no longer the nucleons, 
which are the relevant degrees of freedom. Since there is no strict order parameter for confinement and chiral symmetry breaking, it appears that there is no qualitative difference between low-density nuclear matter and ultra-dense quark matter, such that this transition is allowed to be continuous.  This scenario is realized at zero baryon density, where we do know from first principles that there is a crossover from the hadronic phase to the quark-gluon plasma. At large baryon densities, the situation is more complicated due to the presence of Cooper pairing. Since asymptotically dense and sufficiently cold matter is in the color-flavor locked (CFL) phase \cite{Alford:1998mk,Alford:2007xm}, the actual question is whether one can go continuously from nuclear matter to CFL rather than to unpaired quark matter \cite{Schafer:1998ef,Alford:1999pa,Hatsuda:2006ps,Schmitt:2010pf,Alford:2018mqj,Chatterjee:2018nxe,Cherman:2018jir}. 
Here we will ignore Cooper pairing and address the question of a possible continuity between unpaired, isospin-symmetric nuclear matter and unpaired quark matter, as for instance envisioned 
within a percolation picture \cite{Baym:1979etb,Celik:1980td}.

Combining nuclear matter and quark matter in the same calculation on a microscopic level is very challenging, even if the rigor of a first-principle calculation is given up. However, besides the theoretical motivation just discussed, such studies are highly desired in the context of neutron stars, where for instance mass and radius, but also gravitational wave signals from neutron star mergers, depend on thermodynamic properties of the matter in the core of the star and on the possible existence of a first-order phase transition between nuclear matter and quark matter \cite{Alford:2013aca,Bastian:2018wfl,Christian:2018jyd,Most:2018eaw}. Therefore, various studies exist which combine two distinct models, one with nucleons as degrees of freedom, one with quarks as degree of freedom. They are then either glued together at a first-order phase transition, or a smooth interpolation is introduced between the two phases \cite{Masuda:2012ed,PhysRevD.88.085001,Masuda2016}. 
In either case, these approaches have no predictive power for the nature and the location of the quark-hadron phase transition. Other studies attempt to find a sufficiently general parameterization of the equation of state without relying on a microscopic theory, using microscopic and/or astrophysical input as constraints for the parameterization \cite{Alford:2013aca,Kurkela:2014vha,Tews:2018kmu}. Another strategy is to use a single
model that has either quark degrees of freedom or nucleonic degrees of freedom, but not both. In that case, potential chiral and
deconfinement phase transitions are determined consistently within the given model, but at least one side of the phase transition is then only a very rough approximation to the real world \cite{Schaefer:2007pw,Palhares:2010be,BRATOVIC2013131,Fraga:2018cvr}. An example is the Nambu-Jona-Lasinio (NJL) model \cite{Buballa:2003qv,Ruester:2005jc,Buballa:2016fjh}, which usually does not contain baryons, although it is conceivable to include them \cite{Ishii:1995bu,Alkofer:1994ph,CHRISTOV199691}. 
Our present approach, based on the gauge/gravity duality, is at best a distorted 
version of QCD, in some sense similar to an NJL model, which however {\it does} have a well-defined concept of baryons in a chirally broken phase and of 
chirally symmetric quark matter. Our main results are qualitative and thus at this point we do not attempt to make quantitative predictions for the physics of neutron stars. We do make a step in this direction, however, by fitting the parameters of our holographic model to properties of nuclear matter at saturation
and by calculating the speed of sound for all densities. The results indicate that -- together with future improvements --
our approach can be a useful alternative to the models currently employed for dense matter in neutron stars.
There exist other recent studies that have connected results from the gauge/gravity duality to neutron star physics \cite{Hoyos:2016zke,Annala:2017tqz,Jokela:2018ers}. However, these studies have combined a holographic equation of state for quark matter with a traditional approach for nuclear matter and thus fall in the above category of combining two distinct theories.

\subsection{Method}

We employ the gauge/gravity correspondence ("holography") which allows for a non-pertur\-bative calculation 
of strong-coupling physics via the classical gravity approximation of certain string theories \cite{Maldacena:1997re,Gubser:1998bc,Witten:1998qj}. Specifically, we work with the model developed by Sakai and Sugimoto \cite{Sakai:2004cn,Sakai:2005yt}, who introduced fundamental chiral fermions through $N_f$ D8 and $\overline{\rm D8}$ branes ("flavor branes") into Witten's original model for low-energy QCD based on a background of $N_c$ D4 branes \cite{Witten:1998zw}, see Refs.\ \cite{Peeters:2007ab,Gubser:2009md,Rebhan:2014rxa} for reviews. Here, $N_f$ and $N_c$ are the numbers of flavors and colors, respectively, and the model is usually
evaluated
in the limit $N_f\ll N_c$, which is a necessary condition to neglect effects of the flavor branes on the background geometry.  We shall work in the so-called decompactified 
limit of the model, where the background is given by the deconfined geometry.  
In this limit, the gluon dynamics are decoupled and the dual field theory is expected to resemble the NJL model in certain aspects \cite{Antonyan:2006vw,Davis:2007ka,Preis:2012fh}. The price we have to pay for employing this limit is 
that the rigorous connection to large-$N_c$ QCD is lost -- which in any case only exists in the limit of 
small 't Hooft coupling, which is inaccessible within the gravity approximation. But, in the decompactified limit, 
the chiral phase transition depends nontrivially on the baryon chemical potential even if changes of the background geometry are neglected. This results in a richer
phase structure which may well be closer to
real-world $N_c=3$ QCD, at least with regard to the chiral phase transition. 

Baryons are introduced following the general recipe of the gauge/gravity correspondence, i.e., through branes that 
wrap around an internal subspace of the ten-dimensional geometry and that have $N_c$ string endpoints. In the 
non-supersymmetric Sakai-Sugimoto model, this picture becomes equivalent to instanton solutions of the gauge theory on the flavor branes \cite{Hata:2007mb}. 
Here, "instanton" refers to an object localized in position space and the holographic direction, i.e., they are similar to 
ordinary Yang-Mills instantons with the time direction replaced by the holographic coordinate. The instantons are introduced in the chirally broken phase, where the D8 and $\overline{\rm D8}$ branes are connected, and they carry topological baryon number due to the presence of a Chern-Simons term in the action. This  
is our holographic version of "nuclear matter". In the 
chirally restored phase, the flavor branes are disconnected and baryon number is carried by quarks, represented by strings attached with one end at one of the $N_f$ flavor branes (either D8 or $\overline{\rm D8}$, depending on the chirality of the quarks) and the other end at one of the $N_c$ D4 branes that are responsible for the curved background geometry. This is our holographic version of "quark matter". One of our main results is the observation that the two embeddings of the flavor branes (connected and disconnected) 
can be continuously 
transformed into each other in the presence of instantons. 

Throughout the paper we shall work in the chiral limit. Since the D8 and $\overline{\rm D8}$ branes intercept the D4 branes, quarks are massless in the chirally restored phase.
For discussions about nonzero quark masses in the Sakai-Sugimoto model see 
Refs.\ \cite{Evans:2007jr,Bergman:2007pm,Dhar:2007bz,Aharony:2008an,Hashimoto:2008sr}. In the massless limit, 
chiral symmetry is exact, which suggests a true phase transition between (chirally broken) nuclear matter and (chirally restored) quark matter, either of first or second order. It thus appears 
contradictory to address the question of the continuity between nuclear and quark matter in the chiral limit.
However, our present goal is more modest: we seek to connect chirally broken and chirally restored phases 
in the "order parameter space", along stationary, but not necessarily stable, points of the free energy. This is one step towards an effective potential defined in the entire "order parameter space", which depends on chemical potential and temperature and where nuclear and quark matter correspond to local minima that can be continuously connected whenever they exist. Whether the model allows for a quark-hadron continuity between stable phases can be addressed more comprehensively only after including quark masses.

The main difficulty of our approach is the treatment of the baryonic phase. Single baryons in the 
Sakai-Sugimoto model have been studied extensively in the 
literature, from analytical flat-space approximations to purely numerical studies \cite{Sakai:2004cn,Sakai:2005yt,Hata:2007mb,Seki:2008mu,Cherman:2011ve,Rozali:2013fna}. Here we are interested in a many-baryon system, which requires various approximations. We follow the approach developed in Refs.\ \cite{Ghoroku:2012am,Li:2015uea,Preis:2016fsp,Preis:2016gvf}, which is based on the pioneering work \cite{Bergman:2007wp}, where 
pointlike instantons were considered. Our work is a direct improvement of Ref.\ \cite{Preis:2016fsp}, whose approach we extend by accounting for the 
interactions of the instantons in the bulk.
This is done by using the exact two-instanton solution in flat space, which is a special case of the general Atiyah-Drinfeld-Hitchin-Manin (ADHM) construction
\cite{Atiyah:1978ri}, and which has been discussed previously in the context of the Sakai-Sugimoto model to study 
the nucleon-nucleon interaction \cite{Kim:2008iy,Hashimoto:2009ys}.
As we shall see, the instanton 
interactions are crucial for our main observation.
There are various other, in some aspects complementary, approximations to many-baryon phases in the Sakai-Sugimoto model,
see for instance Refs.\ \cite{Kim:2007vd,Rho:2009ym,Kaplunovsky:2010eh,Kaplunovsky:2012gb,deBoer:2012ij,Bolognesi:2013nja,Kaplunovsky:2015zsa}. One of them is based on a homogeneous ansatz for the gauge fields in the bulk \cite{Rozali:2007rx,Li:2015uea,Elliot-Ripley:2016uwb}, which is 
expected to yield a better approximation for large densities, but which is less transparent from a physical point of view because it is not built from single instantons. 

Our paper is organized as follows. The main point of Sec.\ \ref{sec:setup} is to introduce our ansatz for the non-abelian gauge fields on the flavor branes, based on the single instanton and two-instanton solutions in flat space. This is discussed in Sec.\ \ref{sec:ansatz}
with the technical background deferred to appendix \ref{app:flat}. In Secs.\ \ref{sec:eom} and \ref{sec:mini} we solve the equations of motion and set up the calculation of the stationary points of the free energy. Sec.\ \ref{sec:results} contains our main results. In particular, we discuss the continuity between the chirally broken and chirally symmetric phases in Sec.\ \ref{sec:continuity},
and we compute the speed of sound in Sec.\ \ref{sec:sound}. We give our conclusions in Sec.\ \ref{sec:conclusions}.

\section{Setup}
\label{sec:setup}

\subsection{Action}
\label{sec:action}

The general setup of the model in the given limit has been used in various previous works. The study most relevant to the present paper is Ref.\ \cite{Preis:2016fsp}, whose notation we shall employ. Here we will not go into details regarding the construction of the model itself and refer the reader to the reviews and original works quoted in 
the introduction and, more specifically, to Refs.\ \cite{Li:2015uea,Preis:2016fsp}. The most important difference to 
Ref.\ \cite{Preis:2016fsp} is our ansatz for the non-abelian field strength, which we will explain in detail. 

Our starting 
point is the action for the U($N_f$) gauge fields on the D8 and $\overline{\rm D8}$ branes in the 
deconfined geometry. In our treatment of baryons we shall restrict ourselves to $N_f=2$, i.e., we shall not include hyperons, and thus the non-abelian field strengths will be approximated with the help of 
SU(2) instantons. For consistency we shall use $N_f=2$ also for the quark matter phase, although for our purpose it would be trivial to add a third massless flavor in the chirally symmetric phase. We shall briefly comment on the three-flavor case in Sec.\ \ref{sec:solution}. The action has a Dirac-Born-Infeld (DBI) and a Chern-Simons (CS) contribution,
\be
S= S_{\rm DBI} + S_{\rm CS} \, ,
\ee
with 
\begin{subequations} \label{SDBICS}
\bea
S_{\rm DBI} &=& \frac{{\cal N}}{M_{\rm KK}^3T} \int d^3 x \int_{u_c}^\infty du\, u^{5/2}\non [2ex]
&&\times {\rm str} \sqrt{(1+u^3f_Tx_4'^2+\hat{F}_{u0}^2)\left(1+\frac{F_{ij}^2}{2u^3\lambda_0^2}\right)+\frac{f_TF_{iu}^2}{\lambda_0^2}+\frac{f_T(F_{ij}F_{ku}\epsilon_{ijk})^2}{4u^3\lambda_0^4}} \, , \label{DBI}\\[2ex] 
S_{\rm CS} &=& \frac{3{\cal N}}{2\lambda_0^2M_{\rm KK}^3T} \int d^3 x \int_{u_c}^\infty du\,\hat{a}_0\Tr[F_{ij}F_{ku}]\epsilon_{ijk} \, , \label{CS}
\eea
\end{subequations}
where we have abbreviated 
\be
{\cal N} \equiv \frac{N_c M_{\rm KK}^4\lambda_0^3}{6\pi^2} \, , \qquad \lambda_0\equiv \frac{\lambda}{4\pi} \, , 
\ee
with the 't Hooft coupling $\lambda$. 
For convenience, we have started with a mostly dimensionless formulation, the relation to the corresponding dimensionful quantities
can be found in Refs.\ \cite{Li:2015uea,Preis:2016fsp}. The only dimensionful quantities in the present formulation are in the 
prefactor of the action, the temperature $T$ and the Kaluza-Klein mass $M_{\rm KK}$, which is the inverse radius of the compactified extra dimension with coordinate $x_4$. In the dimensionless coordinates used here, $x_4 \equiv x_4 + 2\pi$.  
The integration over imaginary time has already been performed, yielding the prefactor $1/T$. The remaining integrals 
are taken over position space $\mathbb{R}^3$ and the holographic coordinate $u\in [u_c,\infty]$, where $u_c$ is the 
location of the tip of the connected flavor branes\footnote{The notation "$u_c$" originates from Ref.\ \cite{Bergman:2007wp}, where 
pointlike instantons were used as an approximation, which cause a cusp at the tip of the connected flavor branes, hence the subscript $c$. We keep using this notation even though the embedding is smooth in the presence of instantons with nonzero width.} and the holographic boundary is located at $u=\infty$. The embedding of the flavor branes, given by the function $x_4(u)$ and
by $u_c$ itself will later be determined dynamically, with the boundary condition $x_4(\infty)-x_4(u_c)=\ell/2$, where $\ell$
is the dimensionless asymptotic separation of the D8 and $\overline{\rm D8}$ branes, its dimensionful version given by $L=\ell/M_{\rm KK}$. 
We have denoted derivatives with respect to $u$ by a prime.  The DBI action has been written in terms of the symmetrized 
trace "str", to be taken over the 2-dimensional flavor space. For simplicity, we shall later work with the ordinary trace. 
This is a 
tremendous simplification, and it has been shown for a very similar calculation that the use of the symmetrized 
trace prescription does not make a large quantitative difference in the results \cite{Preis:2016fsp}. The expression in the 
square root is obtained from ${\rm det}(g+2\pi \alpha'{\cal F})$, where $g$ is the induced metric on the flavor branes,
${\cal F}$ is the field strength tensor, and $\alpha'=\ell_s^2$ with the string length $\ell_s$ (which is absorbed in the definition of the dimensionless field strengths). For the explicit form of the metric see for instance Ref.\ \cite{Preis:2016fsp}. It contains the temperature-dependent function 
\be
f_T(u)\equiv 1-\frac{u_T^3}{u^3} \, , \qquad u_T = \left(\frac{4\pi}{3}\right)^2 \frac{t^2}{\ell^2} \, , 
\ee
with the dimensionless temperature $t=T/M_{\rm KK}$. The field strengths are 
decomposed according to ${\rm U(2)}\cong {\rm U(1)}\times {\rm SU(2)}$. Within our ansatz, the only 
nonzero abelian field strength is $\hat{F}_{u0}=i\hat{a}_0'$ with the temporal component of the abelian part of the gauge field
$\hat{a}_0(u)$ (the factor $i$ is needed in the imaginary time formalism). The quark chemical potential enters the action as 
a boundary condition for this abelian gauge field, $\hat{a}_0(\infty) = \mu$. The only nonzero non-abelian field strengths are 
$F_{ij}$ and $F_{iu}$, $i=1,2,3$. They are responsible for the nonzero baryon number through the topological charge provided by
the CS contribution, and their form plays the main role in our calculation. We do not attempt to 
solve the full equations of motion for all gauge fields plus the embedding function in $\vec{x}$ and $u$. (In the action (\ref{SDBICS}) we have already neglected the spatial derivatives of $\hat{a}_0$ and $x_4$.) We shall rather insert an ansatz 
for the square of the non-abelian field strengths, which depends on $\vec{x}$ and $u$, and then average over position space 
such that the spatial integration in the action becomes trivial. Then, we will proceed without further approximation and solve the equations of motion for $\hat{a}_0(u)$ and $x_4(u)$ fully dynamically. 

We will exclusively work with the deconfined geometry, whose metric has been used to derive the DBI action (\ref{DBI}). Since we are interested in the chiral phase transition, we need to work in a regime where the critical temperature of the chiral phase transition is larger than that of the deconfinement transition. Since $T_c^{\rm deconf} = M_{\rm KK}/(2\pi)$ (for all $\mu$ unless backreactions are taken into account) and $T_c^{\rm chiral} \simeq 0.15/L$ (at $\mu=0$) \cite{Aharony:2006da}, this requires $\ell/\pi \lesssim 0.31$ ($\ell=\pi$ is the antipodal limit used in the original works by Sakai and Sugimoto). We shall also assume that the deconfined geometry is preferred down to arbitrarily small temperatures, which is justified in the decompactified limit, where the radius of the $x_4$ circle is taken to infinity (at fixed $L$) such that $M_{\rm KK}$ and thus $T_c^{\rm deconf}$ goes to zero. In this limit, the chiral phase transition is entirely determined by the dynamics on the flavor branes, not by the background geometry.

\subsection{Ansatz for non-abelian field strengths}
\label{sec:ansatz}

With our ansatz for the non-abelian field strengths we attempt to capture as many effects as possible from a many-instanton 
system and at the same time try to keep the expressions as simple as possible to make feasible the full evaluation of the 
equations of motion for the abelian gauge field and the embedding function of the flavor branes. We will construct our 
ansatz from the 1-instanton and the 2-instanton solutions in flat space, which we now discuss. 

\subsubsection{Single instanton}

For large, but not infinite,  't Hooft coupling,
the leading-order expression for a single instanton has been discussed for the confined geometry in Ref.\ \cite{Hata:2007mb} for 
maximally separated flavor branes (= on opposite ends of the $x_4$ circle, $\ell = \pi$) and for non-maximal separation 
in Ref.\ \cite{Seki:2008mu}. Here we need the analogous result for the deconfined geometry from Ref.\ \cite{Preis:2016fsp}, which reads 
\be \label{Fiz0}
F_{iz}^{(1)}F_{iz}^{(1)} = \frac{12(\rho/\gamma)^4}{\gamma^2[x^2+(z/\gamma)^2+(\rho/\gamma)^2]^4} \, ,
\ee
where the superscript "$(1)$" is added to indicate the single-instanton solution, where $x^2 = x_1^2+x_2^2+x_3^2$, and where $z\in [-\infty,\infty]$ is the holographic coordinate along the connected flavor branes, defined by 
\be \label{uz}
u = (u_c^3+u_c z^2)^{1/3} \, , 
\ee
such that $z=0$ corresponds to the tip of the connected branes, $u=u_c$. The shape of 
the instanton is given by 
\be \label{rhogam}
\rho =  \frac{\rho_0u_c^{3/4}}{\lambda^{1/2}}\left[\frac{f_T(u_c)}{\beta_T(u_c)}\right]^{1/4} \sqrt{\alpha_T(u_c)}\, , \qquad \gamma = \frac{3\gamma_0 u_c^{3/2}}{2}\sqrt{\alpha_T(u_c)} \, , 
\ee
where, as Eq.\ (\ref{Fiz0}) shows, $\rho$ can be interpreted as the width in the holographic direction, while $\rho/\gamma$ corresponds to the spatial width. Therefore, $\gamma$ is a 
"deformation parameter", which characterizes the deviation of the instanton from SO(4) symmetry. This deviation has also been
observed in a purely numerical evaluation of a single instanton \cite{Rozali:2013fna}.
In the large-$\lambda$ expansion, $\rho_0=6\sqrt{2\pi/\sqrt{5}}\simeq 10.06$, $\gamma_0=2\sqrt{2/3}\simeq 1.633$ \cite{Preis:2016fsp}, and we have abbreviated
\be
\alpha_T(u_c) \equiv 1-\frac{5u_T^3}{8u_c^3} \, , \qquad \beta_T(u_c) \equiv 1- \frac{u_T^3}{8u_c^3}-\frac{5u_T^6}{16u_c^6} \, ,
\ee
such that at zero temperature $f_T(u_c)=\alpha_T(u_c)=\beta_T(u_c)=1$. 
The solution (\ref{Fiz0}) is the flat-space Belavin-Polyakov-Schwarz-Tyupkin (BPST) instanton \cite{1975PhLB...59...85B}, which is exact for a single instanton in the limit of large $\lambda$ because the width goes to zero for infinite $\lambda$ 
and thus the effect of curved space is negligible. For our many-baryon system this is no longer true, and in principle we would have to construct a (numerical) solution in curved space. For simplicity, however, we shall keep the form (\ref{Fiz0})
as an ansatz. In particular, we shall keep the dependence of $\rho$ and 
$\gamma$ on $u_c$ from Eq.\ (\ref{rhogam}), but treat $\rho_0$ and $\gamma_0$ as free parameters, which can be adjusted to capture effects that 
are known from the numerical solution of a single instanton in curved space and can also be adjusted to reproduce known properties of realistic nuclear matter\footnote{This approach was already suggested 
and explored in Ref.\ \cite{Preis:2016fsp}. However, in this reference, $\rho\propto u_c$, $\gamma\propto u_c^{3/2}$ was used, i.e., the dependence 
of $\rho$ on $u_c$ was chosen differently for simplicity. 
Here we work with the "correct" dependence suggested by 
the lowest-order, single-instanton solution, which leads to a somewhat more difficult calculation, but which turns out to be crucial 
(together with taking into account instanton interactions) for our main observations.}. Since $u_c$ will be determined dynamically for each $\mu$ and $T$, the instanton width in the spatial and holographic directions will depend on density and temperature.

Self-duality of the instanton solution 
implies
\be \label{selfdual}
F_{iz}F_{iz} = \frac{F_{ij}F_{ij}}{2\gamma^2} = -\frac{F_{iz}F_{jk}\epsilon_{ijk}}{2\gamma} \, .
\ee
Note that all field strengths squared are proportional to the $2\times 2$ unit matrix in flavor space. At this point, $\gamma$ appears as a mere rescaling of the holographic coordinate $z$. However, when we insert our ansatz into the DBI action $\gamma$ becomes a nontrivial parameter 
that cannot be eliminated by rescaling $z$. 
The single-instanton solution has topological charge 1, as it should be, 
\be \label{NB}
-\frac{1}{8\pi^2}\int d^3x \int_{-\infty}^\infty dz \, \Tr[F_{ij}^{(1)}F_{kz}^{(1)}]\epsilon_{ijk} = \int_{-\infty}^\infty dz \, q_0(z) = 1 
\, ,
\ee
where, for later convenience,  we have denoted\footnote{In Ref.\ \cite{Preis:2016fsp}, $q_0$ was denoted by $D$.} 
\be
q_0(z) \equiv \frac{3\rho^4}{4(\rho^2+z^2)^{5/2}} \, .
\ee

\subsubsection{Two instantons}

The second ingredient for our ansatz is the flat-space two-instanton solution. If the separation of the two instantons 
is sufficiently large, this solution is well approximated by two single instantons. Let us for now consider SO(4) symmetric instantons, denote the width of both instantons by $\rho$, and let us place one instanton at the origin of the four dimensional $(\vec{x},z)$ space and the other one separated by a distance $\delta$ in the $x_1$ direction. 
Then, the exact flat-space 
solution that we will use is parameterized by $\rho$ and $\delta$  and reads 
\bea \label{calFsq}
F_{iz}^{(2)}F_{iz}^{(2)}&=& \frac{4096 \rho^4}{[(\delta^2 + 4 |x|^2) (\delta^2 + 8 \rho^2 + 4 |x|^2) - 16 \delta^2 x_1^2]^4} \Big\{3 \delta^8
+16 \delta^6 (3 \rho^2 + |x|^2 + 8 x_1^2) \non[2ex]
&& + 
   32 \delta^4 \Big[6 \rho^4 + |x|^4 + 8 x_1^4 + 48 |x|^2 x_1^2  +  4 \rho^2 (|x|^2 + 8 x_1^2)\Big] \non[2ex]
&&+ 256 \delta^2 |x|^2 \Big[|x|^4 + 8 |x|^2 x_1^2 - \rho^2 (|x|^2 - 4 x_1^2) + 768 |x|^8  \Big]  \Big\} \, ,
\eea   
with $|x|^2=x_1^2+x_2^2+x_3^2+z^2$. The parameters $\rho$ and $\delta$ lose their interpretation of width and separation as $\delta$ is decreased and the instantons start to overlap. The derivation of Eq.\ (\ref{calFsq}) is explained in appendix \ref{app:flat}. It is based on the general ADHM construction \cite{Atiyah:1978ri} and its application to the Sakai-Sugimoto model \cite{Kim:2008iy,Hashimoto:2009ys}. For simplicity, we have assumed the instantons to have the same orientation, termed "defensive skyrmions" in Ref.\ \cite{Kim:2008iy}. This renders the subsequent calculation much simpler, but we should keep in mind that in principle the relative orientation of the instantons in the many-instanton system has to be determined by minimizing the free energy. In this sense, the free energy we calculate has to be understood as an upper limit, likely to be reduced by choosing a nontrivial relative orientation.  We introduce the deformation parameter $\gamma$ by the same rescaling as in the single-instanton solution (\ref{Fiz0}): we replace $\rho\to \rho/\gamma$, $z\to z/\gamma$ and divide $F_{iz}^{(2)}F_{iz}^{(2)}$ by $\gamma^2$. Then, we can apply the self-duality (\ref{selfdual}) for the two-instanton solution, and we verify that the topological charge is 2,
\be \label{NB2}
-\frac{1}{8\pi^2}\int d^3x \int_{-\infty}^\infty dz \, \Tr[F_{ij}^{(2)}F_{kz}^{(2)}]\epsilon_{ijk} = \int_{-\infty}^\infty dz \, q_{\rm int}(d,z) = 2  \, .
\ee
Here we have introduced the function 
\bea \label{Phi}
q_{\rm int}(d,z) &=& \frac{3\sqrt{2}\rho^8}{4}\frac{h_1{\cal S}_1+h_2{\cal S}_2}{(a^2+b)^{5/2}b^2} \, ,   
\eea
with the polynomials 
\begin{subequations} \label{p1p2}
\bea
h_1&\equiv& 2\Big[\rho^{10}d^6(4d^4+7d^2-2)^2+\rho^{8}z^2d^2(4d^2-1)^2(4d^6+12d^4+7d^2-2)\non[2ex]
&&+\rho^6z^4(96d^{10}+116d^8-91d^6-14d^4+11d^2-1)+2\rho^4z^6d^2(26d^6-8d^4-23d^2+5)\non[2ex]
&&+4\rho^2z^8d^4(2d^2-3)+4z^{10}d^4\Big] \, , \\[2ex]
h_2 &\equiv& \rho^8d^4(4d^4+7d^2-2)^2+2\rho^6z^2d^2(32d^8+76d^6+15d^4-17d^2+2)\non[2ex]
&&+\rho^4z^4(92d^8+144d^6+5d^4-18d^2+2)+8\rho^2z^6d^2(9d^4+8d^2-2)+28z^8d^4\, ,
\eea
\end{subequations}
and 
\begin{subequations}\label{SSab}
\bea 
a&\equiv& z^2-\rho^2(d^2-1) \, , \qquad b \equiv \rho^2[4d^2(\rho^2+z^2)-\rho^2] \, , \\[2ex]
{\cal S}_1 &\equiv& \sqrt{\frac{-a+\sqrt{a^2+b}}{b}} \, , \qquad 
{\cal S}_2 \equiv \sqrt{a+\sqrt{a^2+b}} \, .
\eea
\end{subequations}
The particular way (\ref{Phi}) of writing the function $q_{\rm int}$ is not only motivated by compactness, but also 
turns out to be useful for integrating and differentiating $q_{\rm int}$. We have also introduced the overlap parameter
\be \label{d}
d\equiv \frac{\delta}{2\rho/\gamma} \, , 
\ee
which is the distance between the instantons normalized by twice their spatial width, such that, if they were spheres with 
radius $\rho/\gamma$, they would overlap for $d<1$.

\subsubsection{Many-instanton approximation}

The ADHM construction in principle provides exact many-instanton solutions, which 
however become increasingly complicated and are much too unwieldy for our purpose. Elements of these 
general solutions have been used in the Sakai-Sugimoto model in Refs.\ \cite{Hashimoto:2009as,Hashimoto:2010ue}, see 
also a brief discussion in Ref.\ \cite{Kaplunovsky:2015zsa}. 
Here we construct a simple approximation for a many-instanton system, taking into account only 2-body interactions. 
To this end, we first define an interaction energy density by subtracting the two individual single-instanton contributions
from the full two-instanton solution,
\be \label{Int}
{\cal I}(\vec{x}_1,z_1,\vec{x}_2,z_2) \equiv (F_{iz}^{(2)})^2(\vec{x}_1,z_1,\vec{x}_2,z_2)- (F_{iz}^{(1)})^2(\vec{x}_1,z_1)-(F_{iz}^{(1)})^2(\vec{x}_2,z_2) \, .
\ee
Via the self-duality (\ref{selfdual}), analogous relations hold for the other two relevant quadratic forms $F_{ij}F_{ij}$ and
$F_{ij}F_{kz}\epsilon_{ijk}$. The two instantons are centered at the points $(\vec{x}_1,z_1)$ and $(\vec{x}_2,z_2)$. 
Each term in Eq.\ (\ref{Int}) also depends on $\vec{x}$ and $z$, which we have not indicated explicitly in the arguments.
The field strength squared for a system of $N_I$ instantons is then approximately constructed as
\be\label{FsqN}
F_{iz}^2 \simeq \sum_n^{N_I} (F_{iz}^{(1)})^2(\vec{x}_n,z_n)+\frac{1}{2}\sum_n^{N_I}\sum_{m\neq n}^{N_I}{\cal I}(\vec{x}_n,z_n,\vec{x}_m,z_m) \, .
\ee
Let us first comment on the holographic location of the instantons. Previous studies have shown that at low baryon density the instantons are located at $z=0$, the tip of the connected flavor branes, while at large density additional 
instanton layers at $|z|>0$ are expected to appear \cite{Kaplunovsky:2012gb,Elliot-Ripley:2015cma,Preis:2016fsp,Elliot-Ripley:2016uwb}. Allowing for  
multiple layers makes our calculation somewhat more complicated, both the numerical evaluation and the following 
equations. Therefore, and since our emphasis is on the new interaction terms, we discuss all modifications that arise 
due to multiple layers in appendix \ref{app:layers}. With the help of this appendix, we have checked that allowing for 
more layers does not change our main conclusion regarding the continuous connection between nuclear and quark matter. 
Here we continue with a single layer, i.e., we center all instantons around the point $z=0$ in the holographic direction. Consequently, $z_n=z_m=0$ in Eq.\ (\ref{FsqN}), and we shall 
omit the arguments referring to the holographic coordinate from now on. Next, we put the instantons on a regular lattice in position 
space and apply the nearest-neighbor approximation. We denote the number of nearest neighbors by $p$ and the lattice vectors
that connect each instanton with its nearest neighbors by $\vec{\delta}_m$, with the nearest neighbor distance $\delta$, i.e., $|\vec{\delta}_1|=\ldots =|\vec{\delta}_p|\equiv \delta$. Then, Eq.\ (\ref{FsqN}) becomes 
\bea \label{Fsum}
F_{iz}^2 &\simeq& \sum_n^{N_I} (F_{iz}^{(1)})^2(\vec{x}_n)+\frac{1}{2}\sum_m^p\sum_{n}^{N_I}{\cal I}(\vec{x}_n,\vec{x}_n+\vec{\delta}_m) \non[2ex]
&\to& N_I\left[(1-p)\left\langle(F_{iz}^{(1)})^2(0)\right\rangle +\frac{p}{2}\left\langle(F_{iz}^{(2)})^2(0,\vec{\delta})\right\rangle\right] \non[2ex]
&=& \frac{2\lambda_0^2 n_I}{3\gamma}\left[(1-p)q_0(z)+\frac{p}{2}q_{\rm int}(d,z)\right] \, .
\eea
In the second line we have employed a further approximation by averaging the field strengths squared over position space, denoted by angular brackets, and used Eq.\ (\ref{Int}). 
This has allowed us to shift every single term of the sum to the origin, $\vec{x}_n\to 0$, with $\vec{\delta}$ connecting an instanton in the origin to one if its nearest neighbors, say along the $x_1$ direction. 
We have used the spatial integrals 
for the one-instanton and two-instanton solutions from Eqs.\ (\ref{NB}) and (\ref{NB2}), respectively. Moreover, we have
introduced the dimensionless instanton density (per flavor)
\be
n_I = \frac{48\pi^4 N_I}{\lambda^2 v} \, ,
\ee
where $v$ is the dimensionless three-volume of the system, related to the 
dimensionful volume $V$ by $v=VM_{\rm KK}^3$. The instanton distance $d$ in the last line of Eq.\ (\ref{Fsum}) is of course related to the instanton density. This relation 
depends on the lattice structure. We write the volume of a Wigner-Seitz cell as 
\be \label{s}
\frac{v}{N_{I}}= \frac{\delta^3}{r} \, ,
\ee
such that $r$ is the inverse Wigner-Seitz volume in units of the nearest-neighbor distance $\delta$.
Consequently,
with Eq.\ (\ref{d}) we can express $d$ as 
\be \label{d1}
d  = \frac{\gamma}{\rho}\left(\frac{6\pi^4 r}{\lambda^2 n_I}\right)^{1/3} \, .
\ee
The parameters $p$ and $r$ carry all information about the lattice structure in our approximation. For a cubic, a body centered cubic, and a face centered cubic crystal we have $(p,r) = (6,1), (8,3\sqrt{3}/4), (12,\sqrt{2})$, respectively. By setting the number of nearest neighbors $p$ 
to zero the result becomes independent of $r$ and we recover the non-interacting 
approximation of Ref.\ \cite{Preis:2016fsp}. Replacing the field strengths squared 
by their spatial average is an enormous simplification since now the equations of motion 
will be ordinary differential equations, only depending on the holographic coordinate $u$. A more refined approximation would be to average not over the squared field strengths, but over the entire square root of the DBI action (\ref{DBI}) before solving the equations of motion. Although still resulting in ordinary differential equations, this procedure would be much more complicated since 
there is no simple analytic form for the spatial integral over the DBI Lagrangian, even if the 
above analytic ansatz for the non-abelian field strengths is used.

\subsection{Solving the equations of motion}
\label{sec:eom}

We now insert our ansatz (\ref{Fsum}) together with the corresponding expressions for $F_{ij}^2$, $F_{ij}F_{kz}\epsilon_{ijk}$ into 
the action (\ref{SDBICS}). As already mentioned, instead of the symmetrized trace we take the standard trace for simplicity. 
Since all field strengths squared are diagonal in flavor space, this simply gives an overall factor $N_f$, and we can write 
\be \label{S1}
S = {\cal N} N_f \frac{V}{T} \int_{u_c}^\infty du\, {\cal L} \, , 
\ee
with the Lagrangian 
\be \label{Lag}
{\cal L} = u^{5/2}\sqrt{(1+u^3f_Tx_4'^2-\hat{a}_0'^2+g_1)(1+g_2)}-\hat{a}_0 n_I q(u) \, ,
\ee
where
\be
g_1 \equiv \frac{f_T n_I}{3\gamma}\frac{\partial z}{\partial u} q(u) \, , \qquad g_2 \equiv \frac{\gamma n_I}{3u^3}\frac{\partial u}{\partial z} q(u) \, ,
\ee
with 
\be
q(u) \equiv 2\frac{\partial z}{\partial u} \left[(1-p)q_0(z)+\frac{p}{2}q_{\rm int}(d,z)\right] \, , \qquad \int_{u_c}^\infty du\, q(u)  = 1 \, .
\ee
(Here we are switching back and forth between the coordinates $u$ and $z$, whose relation is given in Eq.\ (\ref{uz}).) 
We have brought the Lagrangian into the same form as in Ref.\ \cite{Preis:2016fsp}, with all effects of the instanton interaction absorbed in the function $q(u)$. Therefore, for the solutions of the equations 
of motion $\hat{a}_0(u)$ and $x_4(u)$, we can simply follow this reference. With the integration constant $k$ we 
obtain 
\bea \label{a0x4}
\hat{a}_0'&=& \frac{n_IQ}{u^{5/2}} \zeta\, , \qquad x_4'=\frac{k}{u^{11/2}f_T}\zeta \, ,
\eea
where we have abbreviated 
\be \label{zeta}
\zeta\equiv \frac{\sqrt{1+g_1}}{\sqrt{1+g_2-\frac{k^2}{u^8f_T}+\frac{(n_IQ)^2}{u^5}}} = \frac{\sqrt{1+u^3f_Tx_4'^2-\hat{a}_0'^2+g_1}}{\sqrt{1+g_2}}\, ,
\ee
and
\bea \label{Q}
Q(u) &\equiv& \int_{u_c}^u du'\, q(u') = (1-p)Q_0(z)+ \frac{p}{2}Q_{\rm int}(d,z) \, ,
\eea
where
\begin{subequations}
\bea
Q_0(z) &\equiv& \int_{-z}^z dz'\,q_0(z') = \frac{(3\rho^2 + 2z^2)|z|}{2(z^2+\rho^2)^{3/2}} \, , \\[2ex]
Q_{\rm int}(d,z) &\equiv&  \int_{-z}^z dz'\,q_{\rm int}(z') = \frac{\sqrt{2}\rho^2}{2}\frac{H_1 {\cal S}_1+H_2{\cal S}_2}{(a^2+b)^{3/2}b} \, , \label{Psi}
\eea
\end{subequations}
 with the polynomials 
\begin{subequations} \label{q1q2}
\bea
H_1&=&\rho^2 z \Big[6  \rho^6 d^4 (4 d^4 +7 d^2 -2 )  +  \rho^4 z^2 (16 d^8 + 80 d^6 + 36 d^4- 16 d^2+1) \non[2ex]
&& + 2  \rho^2 z^4 d^2 (16 d^4+ 28 d^2 -5 ) + 16  z^6 d^4\Big] \, , \\[2ex]
H_2&=& z\Big[3  \rho^6 d^2 (4 d^4 + 7 d^2-2  ) +\rho^4 z^2 (8 d^6+ 40 d^4 + 17 d^2-5   ) \non[2ex]
&&+ 2\rho^2 z^4 (8 d^4 + 14 d^2 -1) + 8 z^6 d^2  \Big] \, ,
\eea
\end{subequations}
and ${\cal S}_1$, ${\cal S}_2$, $a$, $b$ defined in Eq.\ (\ref{SSab}). At first sight, 
finding an analytic expression for the integral in Eq.\ (\ref{Psi}) seems hopeless, due to the 
complicated form of $q_{\rm int}$. As explained in appendix \ref{app:deriv}, however, an analytical form can be found  
from the observation that taking successive derivatives does not change the structure of the function, only the exponents in the denominator and the explicit form of the polynomials that multiply ${\cal S}_1$ and ${\cal S}_2$. 

The solution (\ref{a0x4}) includes the case of a trivial embedding function, $x_4'=k=0$, which corresponds to the chirally symmetric quark matter phase. It also includes the mesonic phase, which has vanishing baryon number, $n_I=0$, and thus a constant field $\hat{a}_0(u)=\mu$. In the baryonic phase both $k$ and $n_I$ are nonvanishing. For all cases, the asymptotic behavior at the holographic boundary $u\to \infty$ can be written as 
\be
\hat{a}_0'(u) = \frac{n_I}{u^{5/2}} +\ldots \, , \qquad x_4'(u) = \frac{k}{u^{11/2}} + \ldots \, . 
\ee
This expansion follows from Eq.\ (\ref{a0x4}) and $Q=1+{\cal O}(u^{-6})$, $\zeta = 1+{\cal O}(u^{-5})$ and confirms the interpretation of $n_I$ as the baryon number density.

\subsection{Stationary points of the free energy}
\label{sec:mini}

According to the on-shell action (\ref{S1}) the free energy is ${\cal N}N_f\Omega$ with 
\be \label{Omega}
\Omega = \int_{u_c}^\infty du\, {\cal L}  \, .
\ee
Here, the Lagrangian is obtained by re-inserting the solutions (\ref{a0x4}) into Eq.\ (\ref{Lag}).
As written, the free energy
is divergent. We can easily deal with this divergence by introducing a cutoff
for the upper boundary of the $u$ integral. This introduces a cutoff dependent term that turns out to be independent of 
chemical potential and temperature. We can therefore simply subtract this vacuum term since it does not affect the physics. 

The free energy depends on various parameters. Firstly, there are the model parameters  
$\lambda$, $M_{\rm KK}$, and $\ell$. 
Secondly, there are the quantities $k$, $n_I$, $u_c$, $\rho$, $\gamma$, which, in principle, have to be determined dynamically. However, as shown in Ref.\ \cite{Preis:2016fsp}, this leads to pointlike instantons at the
baryon onset (which, unphysically, is of second order), while we know that at finite $\lambda$ 
the instantons should have a nonzero width. We have checked that this observation remains unchanged in the present approach and thus we shall use the form of $\rho$ and $\gamma$ from Eq.\ (\ref{rhogam}) and treat $\rho_0$ and $\gamma_0$ as parameters, increasing the number of model parameters from 3 to 5 \cite{Preis:2016fsp}. This can be understood as an attempt to capture nontrivial effects, for instance known from other approximations and numerical studies, for which our original ansatz is too simplistic. It can also be understood as 
a phenomenological approach, since it gives us additional freedom to reproduce properties of real-world nuclear matter. 
Within this approach, it remains to find the stationary points of $\Omega$ with respect to $k$, $n_I$, and $u_c$. The three stationarity equations
\be
\frac{\partial \Omega}{\partial k} = \frac{\partial \Omega}{\partial n_I} = \frac{\partial \Omega}{\partial u_c} = 0
\ee
can be written more explicitly as
\begin{subequations} \allowdisplaybreaks \label{minimOm}
\bea
\frac{\ell}{2} &=& \int_{u_c}^\infty du\, x_4' \, , \label{dk} \\[2ex]
\mu n_I &=&\int_{u_c}^\infty du \,u^{5/2}\left[\frac{g_1\zeta^{-1}+g_2\zeta}{2q}\left(q-\frac{d}{3}\frac{\partial q}{\partial d}\right) + \frac{\zeta n_I^2 Q}{u^5}\left(Q -\frac{d}{3}\frac{\partial Q}{\partial d}\right)\right] 
\, , \label{dnI} \\[2ex]
 st &=& 2u_c^{7/2}+\int_{u_c}^\infty du \, u^{5/2} \left\{7 - \zeta\left[7(1+g_2)+2\frac{(n_I Q)^2}{u^5}+\frac{k^2}{u^8f_T}\right]\right.
\non[2ex]
&&\left.+\frac{g_1\zeta^{-1}+g_2\zeta}{2q}\left(5q-\frac{3d}{2}\frac{\partial q}{\partial d}
 +\frac{\rho}{2}\frac{\partial q}{\partial \rho}\right)-\frac{\zeta n_I^2Q}{u^5}\left(\frac{3d}{2}\frac{\partial Q}{\partial d}-\frac{\rho}{2} \frac{\partial Q}{\partial \rho}\right)\right\}   \, , \label{duc0}
\eea
\end{subequations}
where the derivatives with respect to $d$ are taken at fixed $\rho$ and vice versa.
The first two equations, the derivatives with respect to $k$ and $n_I$, are straightforwardly derived. 
The derivative with respect to $u_c$ is more complicated, and we explain the derivation in appendix \ref{app:mini}. We have denoted the dimensionless entropy density by $s$, 
\be \label{entropy}
s = -\frac{\partial \Omega}{\partial t} = \frac{3u_T^3}{t}\int_{u_c}^\infty \frac{du}{u^{1/2}f_T}\left(g_1\zeta^{-1}+\frac{k^2}{u^8f_T}\zeta \right) \, . 
\ee
Note that $s$ is computed solely from the derivative with respect to the explicit dependence on the dimensionless temperature $t$ because the implicit dependence through $k$, $n_I$, $u_c$ vanishes at the stationary point.

We see from Eq.\ (\ref{dk}) that stationarity with respect to $k$ is equivalent to the boundary condition 
$x_4(\infty)-x_4(u_c)=\ell/2$, given by the fixed asymptotic separation of the flavor branes. Therefore, 
Eq.\ (\ref{dk}) should be viewed as a constraint, it is not a minimization of the free energy. 
This constraint restricts the parameter space to a two-dimensional surface in $(k,n_I,u_c)$ space. 
Local minima of the potential are points that fulfill Eqs.\ (\ref{minimOm}) and in whose vicinity the potential increases in all directions on this two-dimensional surface (and not necessarily in all directions in the  three-dimensional $(k,n_I,u_c)$ space).

\subsection{Choice of parameters}
\label{sec:parameters}

The stationarity equations (\ref{minimOm}) have to be solved simultaneously for $k$, $n_I$, $u_c$ for given chemical potential $\mu$ and temperature $t$. In the remainder of the paper we shall set $t=0$. Moreover, we shall work with the cubic lattice, $p=6$, $r=1$. We have checked that other lattice structures slightly change our results 
quantitatively, but not qualitatively. It remains to fix the model parameters $\lambda$, $M_{\rm KK}$, $\ell$, $\rho_0$, $\gamma_0$. Here we do not attempt to study this parameter space systematically, as it 
was done in Ref.\ \cite{Preis:2016fsp} within a simpler version of the model. 
 
We shall rather work with a single parameter set, which we determine by requiring the model to reproduce the basic properties of nuclear matter 
at saturation (a first step in this direction was made in Ref.\ \cite{Preis:2016gvf}). This is, firstly, a useful preparation  
for future extensions and applications of this model. For instance, if the equation of state is calculated and used to describe dense matter inside neutron  stars, it is important to anchor the approach to known low-density properties. And, secondly, 
the fact that it is possible to reproduce properties of real-world nuclear matter can be viewed as a phenomenological 
validation of the model. We will reproduce 4 properties with 5 free parameters, and thus the fact that the matching works might not seem very impressive. But, due to the complicated, highly nonlinear structure of the stationarity equations it is a priori not obvious that the equations set by the constraints do have a solution.  

The physical constraints we impose are as follows.  We require the vacuum mass of the nucleon to be $m_N = 939\, {\rm MeV}$. The remaining conditions are properties of infinite, isospin-symmetric, zero-temperature nuclear matter at saturation: the binding energy $E_B=-16.3\, {\rm MeV}$, the saturation density $n_0 = 0.153\, {\rm fm}^{-3}$, and the incompressibility, for which only a certain range is known,  $K\simeq(200-300)\, {\rm MeV}$. Relating these quantities to our model gives the following four conditions, 
\begin{subequations} \label{4conds}
\bea
m_N+E_B &=& \frac{\lambda N_cM_{\rm KK}}{4\pi} \,\mu_0 \, , \label{binding}\\[2ex]
m_N &=& \frac{\lambda N_cM_{\rm KK}}{4\pi} \, m_0 \, , \label{mN} \\[2ex]
n_0&=&\frac{\lambda^2 M_{\rm KK}^3}{48\pi^4} \,n_I^0 \, , \label{n0} \\[2ex]
K&=&\frac{\lambda N_c M_{\rm KK}}{4\pi}\,\kappa  \label{comp}\, .
\eea
\end{subequations}
The right-hand sides depend on the parameters of the model, through the dimensionless quantities $\mu_0$, $m_0$, $n_I^0$, $\kappa$, which are functions of $\rho_0$, $\gamma_0$, $\lambda$, and $\ell$. Here, $\mu_0$ is the quark chemical potential at the baryon onset, and $n_I^0$ is the baryon density just above the onset. To relate $\mu_0$ and $n_I^0$ to their dimensionful counterparts, we have used Eqs.\ (2.44) and (2.45) of Ref.\ \cite{Preis:2016fsp}. Moreover, $m_0$ is the constituent quark mass in the vacuum, such that $N_cm_0$ is the vacuum 
mass of the baryon. In our approach, it can be computed from the chemical potential in the limit of vanishing baryon density, $n_I\to 0$. Since in this limit the instantons are infinitely far away from each other we can obviously use the non-interacting limit to compute the vacuum mass. We find that the solution of Eqs.\ (\ref{minimOm}) for $n_I\to 0$ 
with $\mu$ approaching a nonzero value is (for zero temperature)
\begin{subequations}\allowdisplaybreaks
\bea 
u_c &=& \frac{16\pi}{\ell^2}\left[\frac{\Gamma\left(\frac{9}{16}\right)}{\Gamma\left(\frac{1}{16}\right)}\right]^2 \, ,\label{ucMes}  \\[2ex]
k&=&u_c^4 \, , \\[2ex]
m_0&\equiv & \mu = \frac{3u_c^3\rho_0^4}{8\gamma_0\lambda^2}\int_{u_c}^\infty du\,u^{5/2} \frac{u^8-u_c^8+\gamma_0^2 u_c^4u(u^3-u_c^3)}{(u^3-u_c^3)\sqrt{u^8-u_c^8}\left(u^3-u_c^3+\frac{\rho_0^2u_c^{5/2}}{\lambda}\right)^{5/2}}
\, .\label{M0}
\eea
\end{subequations}
We mention already here that there is a second solution for $n_I\to 0$, where $\mu$ does go to zero as well. This 
solution will play  crucial role later, see Eq.\ (\ref{smallMu}). Finally, $\kappa$ is the dimensionless version of the incompressibility at saturation,
\be \label{kappa}
\kappa = - 9 \left.\frac{\partial \Omega}{\partial n_I}\right|_{n_I=n_I^0} \, .
\ee
This derivative with respect to the baryon density is taken 
at fixed temperature,
but with $\mu$, $k$, $u_c$ being functions of 
$n_I$. In contrast, the same derivative, used above for minimizing the free energy, was taken at fixed temperature {\it and} fixed $\mu$, $k$, $u_c$. We compute the derivative in (\ref{kappa}) purely numerically. 

The 4 conditions (\ref{4conds}) are used to fix $\rho_0$, $\gamma_0$, $\lambda/\ell$, and $\ell/M_{\rm KK}=L$. This matching is not unique because $K$ is not known exactly and because it is conceivable that different, disconnected regions in parameter space satisfy the physical constraints. It is beyond the scope of this paper to present a systematic analysis of the multi-dimensional parameter space. The parameter set that will be used throughout the 
rest of the paper is   
\be \label{parachoice}
\rho_0 = 4.3497 \, , \qquad \gamma_0 = 3.7569 \, , \qquad \frac{\lambda}{\ell}=15.061 \, , \qquad \frac{\ell}{\pi}=\frac{M_{\rm KK}}{3185\, {\rm MeV}} \, .
\ee
It can be checked numerically that these values satisfy Eqs.\ (\ref{4conds}): when solving 
the stationarity equations (\ref{minimOm}) it is useful to eliminate $\ell$ completely from the equations by an appropriate rescaling of all 
quantities. After that rescaling, the stationarity equations yield, using the values (\ref{parachoice}), $\mu_0\simeq0.2531\,\ell^{-2}$, $m_0\simeq 0.2576\, \ell^{-2}$, $n_I^0 \simeq 0.02325\, \ell^{-5}$, $\kappa\simeq 0.06\,  \ell^{-2}$. It is then easy to check that these values satisfy the physical conditions at saturation; in particular $K\simeq 220\, {\rm MeV}$, which is within the experimentally known range \cite{Blaizot:1995zz,Youngblood:2004fe}, possibly somewhat smaller \cite{Stone:2014wza}. The predicted value for $L$ from Eq.\ (\ref{parachoice}) can 
be used to compute the critical temperature for the chiral transition at zero chemical potential. 
Equating the free energies of the mesonic phase and the quark matter phase -- for instance using 
appendix B of Ref.\ \cite{Li:2015uea} -- yields a first-order phase transition with $T_c^{\rm chiral}\simeq 0.15384/L$, from which we obtain $T_c^{\rm chiral}\simeq 156\, {\rm MeV}$. This is very close to the transition temperature for the chiral crossover in real-world QCD. We should keep in mind, however, that we work in the chiral limit with two quark flavors. Since lattice simulations are not available in this limit, the exact value and even the order of the transition is not rigorously known. It is nevertheless worth pointing out that from fixing our parameters to cold nuclear matter, we obtain a prediction for the finite-temperature phase transition which can in principle be compared to QCD.

Besides the absolute value of $L$ let us also briefly comment on the relation 
between $\ell$ and $M_{\rm KK}$ in Eq.\ (\ref{parachoice}). This relation suggests that by choosing $M_{\rm KK}$
arbitrarily any value of $\ell$ can be obtained. However, $\ell$ is constrained because the asymptotic separation of the flavor branes can obviously not be larger than half of the circumference of the $x_4$ circle, and hence $\ell<\pi$ is an absolute limit. As discussed at the end of Sec.\ \ref{sec:action}, a more restrictive condition comes from the observation that there is a chirally broken phase in the deconfined geometry only if $\ell/\pi<0.30768$.  Interestingly, at this upper limit, we have $\lambda\simeq 14.6$ and $M_{\rm KK}\simeq 980\, {\rm MeV}$, which is close to the original fits in the confined geometry, which were obtained using completely different constraints based on the rho meson mass and the pion decay constant \cite{Sakai:2004cn}.   
The decompactified limit, which allows us to apply the deconfined geometry to arbitrarily small temperatures, implies $\ell\ll \pi$. It is a matter of interpretation whether this stronger constraint should be obeyed since one might interpret the decompactified limit as a separate model, allowing for extrapolations of $\ell/\pi$ beyond the originally allowed regime. Nevertheless, it is reassuring that $\ell\ll\pi$ does not contradict our physical constraints.

\section{Results}
\label{sec:results}

\subsection{Solution of the stationarity equations}
\label{sec:solution}

Having fixed all parameters, we can solve the stationarity equations and compute the free energy. This has to be done numerically in general, but in one instance we have found an analytical expansion. This is the limit of small $\mu$, where the solution to Eqs.\ (\ref{minimOm}) is
\be \label{smallMu}
u_c = \frac{ 2^4\cdot3^2}{\phi^2}\,\mu^2 + \ldots \, , \qquad k = \frac{2^{16}\cdot3^{11} \ell}{\phi^9}\, \mu^9 + \ldots\, , \qquad n_I = \frac{2^{15}\cdot3^7 }{\phi^7} \mu^6 + \ldots\, ,
\ee
where we have abbreviated 
\be
\phi\equiv \frac{\rho_0^2}{\lambda\gamma_0} \, .
\ee
With Eq.\ (\ref{d}), this solution yields in particular $d\propto \mu^{-1/2}$, 
i.e., the ratio of the distance over the width of the instantons becomes infinitely large as $\mu$ goes to zero. As a consequence, the instantons are essentially non-interacting in this limit, and Eq.\ (\ref{smallMu}) does not depend on the interaction terms\footnote{The solution (\ref{smallMu}) does not exist if $\rho\propto u_c$ instead of $\rho\propto u_c^{3/4}$, which explains why this solution was not found in 
Ref.\ \cite{Preis:2016fsp}.}. 

\begin{figure}[tbp]
\centering 
\hbox{\includegraphics[width=.5\textwidth]{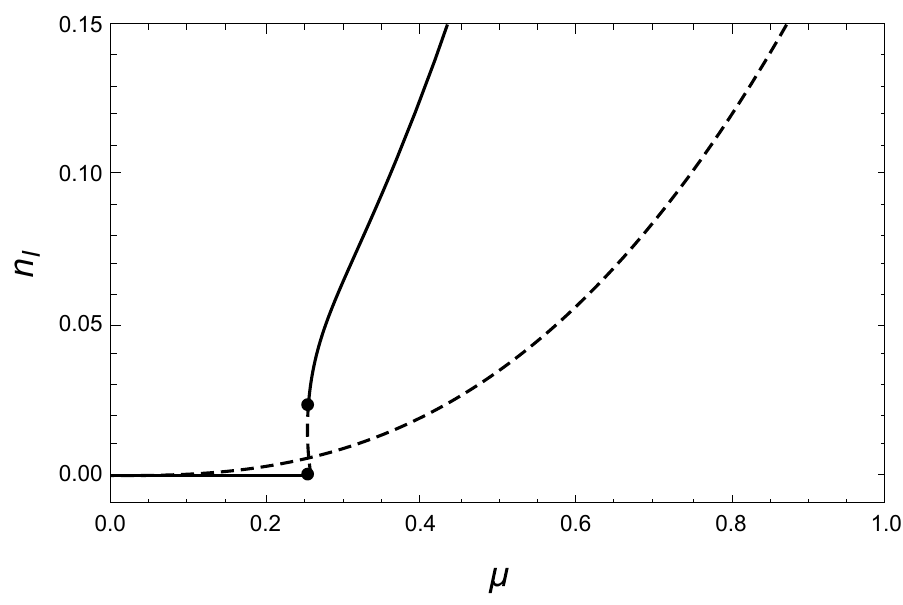}\includegraphics[width=.5\textwidth]{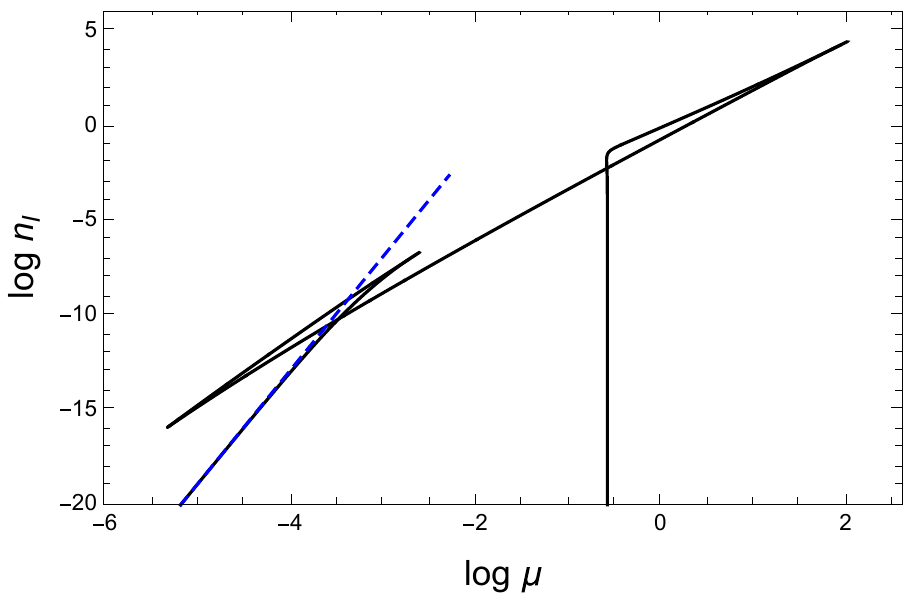}}

\hbox{\includegraphics[width=.5\textwidth]{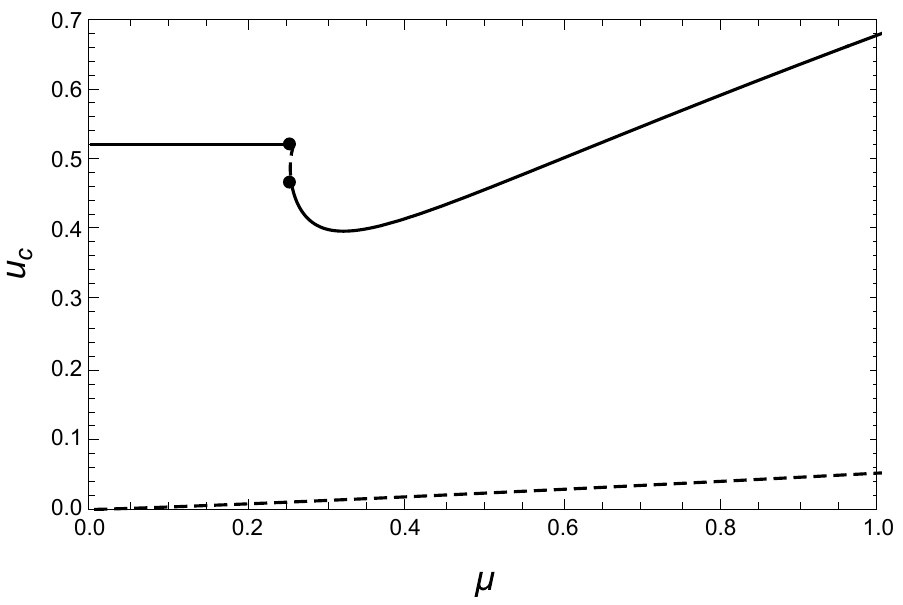}\includegraphics[width=.5\textwidth]{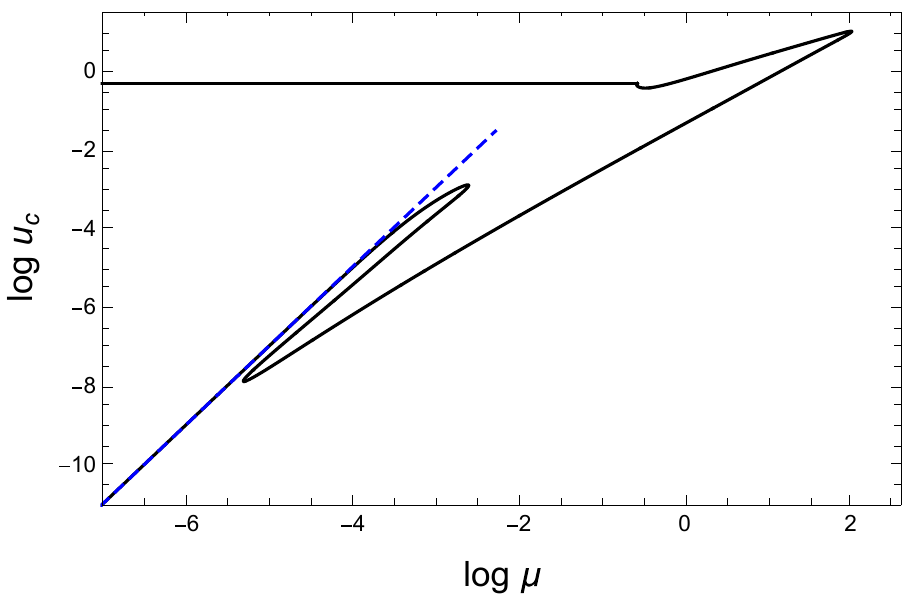}}
\caption{\label{fig:nIuc} Dimensionless baryon density $n_I$ (upper panels) and location of the tip of the connected flavor branes $u_c$ (lower panels) as a function of the dimensionless chemical potential $\mu$ at zero temperature. In the left panels, solid lines are the stable branches, dashed lines are metastable or unstable. The right panels show the same quantities in double-logarithmic plots, now all branches shown as solid lines, together with the analytical result for small $\mu$
(blue dashed). 
} 
\end{figure}

We show the full numerical solution in Fig.\ \ref{fig:nIuc}. In this figure, we have plotted $n_I$ (upper panels) and 
$u_c$ (lower panels)\footnote{Here and in all following figures, the axes labels have to be understood to include appropriate powers of $\ell$, i.e., $n_I$ stands for $n_I\ell^5$, $\mu$ for $\mu \ell^2$, $u_c$ for $u_c \ell^2$, and $\Omega$ for $\Omega\ell^7$; the overlap parameter $d$ does not scale with $\ell$.}. The third variable $k$ is not plotted because 
it shows a qualitatively identical behavior as $u_c$, although there is no simple analytical relation between them. Each quantity is shown in a linear plot (left) and a double-logarithmic plot (right) to make all important features of the solution visible. In the left plots, solid lines corresponds to the stable solution, dashed lines indicate
metastable and unstable branches of the solution. 
Fig.\ \ref{fig:nIuc} also includes the result from the mesonic phase, which has zero baryon number, $n_I=0$ and where
$u_c$ is constant in $\mu$ and assumes the value (\ref{ucMes}). 
We see that there is a first-order baryon onset, which occurs by construction due to our 
matching of the parameters to real-world nuclear matter at saturation. The double-logarithmic plots show that the two branches in the left plots are continuously connected. As a check, we have also included the analytical result (\ref{smallMu}) to confirm that it is in exact agreement with the full 
result at (very) small $\mu$. Moreover, we see that at small, but not too small, chemical potentials, $n_I$ and $u_c$ become three-valued in the baryonic phase. This multivaluedness is not visible in the linear plots. We do not have any particular interpretation for this behavior. It does not play any role for the ground state since in this regime the mesonic phase has a smaller free energy than any of the three baryonic solutions. We shall thus not further discuss this feature, although it is prominently visible in the double-logarithmic plots. 

\begin{figure}[tbp]
\centering 
\hbox{\includegraphics[width=.5\textwidth]{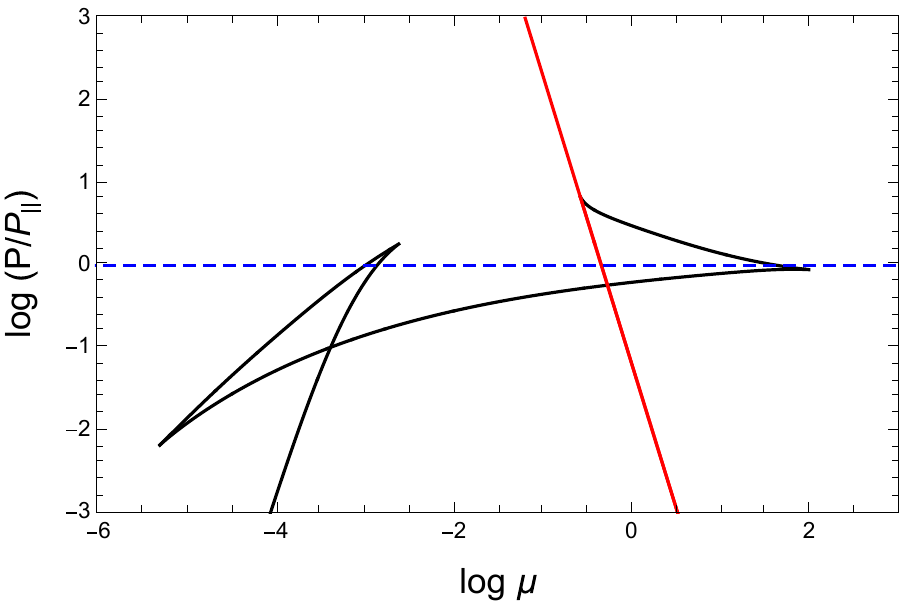}\includegraphics[width=.5\textwidth]{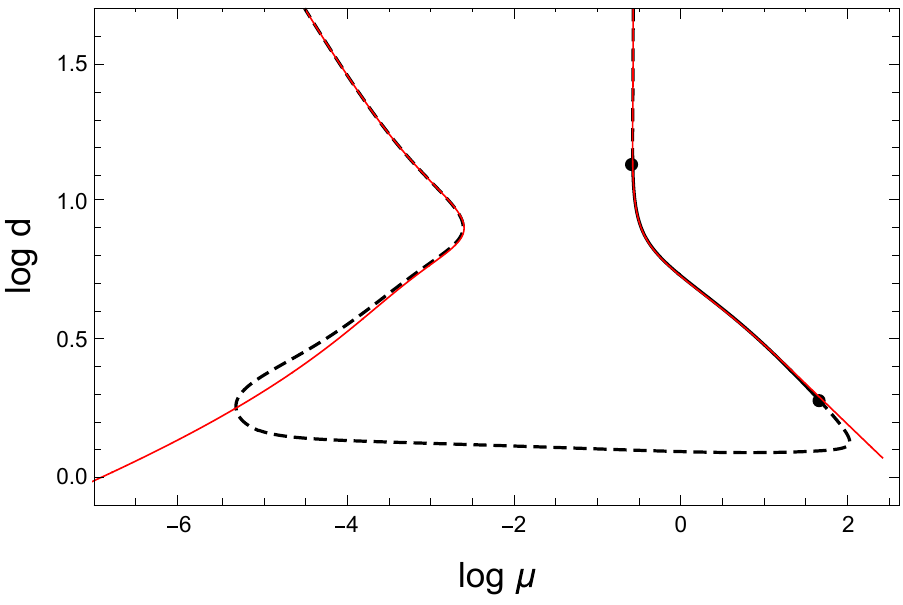}}
\caption{\label{fig:dP} {\it Left panel:} Pressure of the baryonic phase (black solid), mesonic phase (straight red solid line), and chirally symmetric phase (blue dashed horizontal line), all normalized to the pressure of the chirally symmetric phase $P_{||}$. {\it Right panel:} instanton distance over instanton width $d$ as a function of $\mu$ for the baryonic phase. The (black) solid 
segment corresponds to the stable phase, while the dashed segments are unstable or metastable. The thin (red) curves 
are the results with non-interacting instantons. 
} 
\end{figure}

In the left panel of Fig.\ \ref{fig:dP} we show the pressure $P=-\Omega$ for the mesonic phase (straight red line) and the baryonic phase (solid black curve), both divided by the pressure of the quark phase $P_{||}$, which is represented by the (blue) dashed horizontal line. Here, the zero-temperature results for the mesonic and quark phases are (see for instance Ref.\ \cite{Li:2015uea}),
\begin{subequations}
\bea
P_{\cup} &=& \frac{2^{15}\pi^4}{7\ell^7}\frac{\Gamma\left(\frac{15}{16}\right)\tan\frac{\pi}{16}}{\Gamma\left(\frac{7}{16}\right)}\left[\frac{\Gamma\left(\frac{9}{16}\right)}{\Gamma\left(\frac{1}{16}\right)}\right]^7 \, , \label{OmM}\\[2ex]
P_{||} &=& \frac{2}{7}\left[\frac{\sqrt{\pi}}{\Gamma\left(\frac{3}{10}\right)\Gamma\left(\frac{6}{5}\right)}\right]^{5/2}\mu^{7/2} \, , \label{OmQ}
\eea
\end{subequations}
where the subscripts indicate the shape of the flavor branes of the two phases. 
The favored state for a given $\mu$ is the one with the largest value of $P/P_{||}$. We see that for small $\mu$ the mesonic phase (the vacuum) is preferred, 
until at $\mu_0\simeq 0.253/\ell^2$ there is a first-order phase transition to the baryonic phase.  This first-order
phase transition is weak in the sense that the chemical potential at the transition is not much smaller than the vacuum mass of the nucleon, as required from the matching to QCD. Therefore, in the double-logarithmic plot, the multivaluedness of the baryonic pressure around this first-order phase transition is not visible. At $\mu\simeq 42.7/\ell^2$ there is a
first-order phase transition to the chirally symmetric phase. In physical units, this 
corresponds to a critical chemical potential of about 160 GeV or, using the value for the dimensionless baryon density at 
the transition $n_I\simeq 3780/\ell^5$, to more than $10^5$ times saturation density. This is an enormously large value, for instance compared to the chiral transition suggested by phenomenological models, but also because at these ultra-high 
densities we are approaching the regime where perturbative QCD is expected to be valid. If 
two-layer or multi-layer instanton solutions are taken into account, which decrease the free energy of the baryonic phase \cite{Preis:2016fsp}, the chiral phase transition will be even further shifted to 
larger densities. On the other hand, it has been shown, at least without instanton interactions \cite{Preis:2016fsp}, that the chiral phase transition can be arbitrarily close to the baryon onset 
for certain choices of the parameters $\gamma_0$ and $\rho_0$, which then, however, lead to unrealistically large binding energies of nuclear matter \cite{Preis:2016gvf}. In any case, we should take the large value of the critical chemical potential with care for various reasons. 
Firstly, we have neglected backreactions on the background geometry, which can be expected to become large at large baryon densities. Even if our approximation is taken seriously, the large value may be 
a consequence of the large-$N_c$ limit that is inherent in our calculation and/or the lack of asymptotic freedom of the model. We should also recall that for consistency we work with two flavors also in the chirally symmetric phase. In reality, however, {\it strange} quark matter is the ground state at high densities.  (Even 
more quark flavors appear if the chemical potential is sufficiently large, but, eventually having in mind neutron star applications, let us ignore them.) In our approximation of massless quarks, an additional quark flavor in the chirally symmetric phase simply changes the prefactor of the free energy from $N_f=2$ to $N_f=3$. Since the free energy is negative, this reduces the free energy, i.e., makes strange quark matter more favorable, as it should be, and we find a critical chemical potential of about 30 GeV, still large, but reduced by about a factor 5 compared to two-flavor quark matter. Since here we are mostly interested in qualitative results, we leave a more systematic discussion of the 
chiral phase transition for the future.

In the right panel of Fig.\ \ref{fig:dP} we show the instanton overlap parameter $d$. The (black) solid curve, bounded by the two dots, is the stable baryonic segment, from the baryon onset (large $d$) to the chiral transition (smaller $d$), while the dashed
segments are unstable or metastable. The result shows that the instantons "refuse" to overlap strongly. Even at the highest densities we have $d>1$, because, as the instantons are squeezed, they become smaller. As a consequence, the non-interacting solution is a very good approximation for large parts of the curve. This solution, obtained in our approximation 
by setting the number of nearest neighbors to zero, $p=0$, is shown by the two thin (red) curves. The right branch of the non-interacting result ends at large $\mu$, we have not found any solutions beyond this point. Since the numerics get increasingly difficult at these large chemical potentials we cannot exclude that this branch continues beyond the point where we have stopped. On the other hand, we were able to continue the left branch to much smaller values of $\mu$ than shown in the plot. Therefore, if instanton interactions are neglected these two branches appear to be disconnected. 
Only in the presence of interactions all baryonic solutions we have found are continuously connected. And, they
are also continuously connected to the chirally symmetric phase, as we shall discuss now.

\subsection{Connecting nuclear matter with quark matter}
\label{sec:continuity}

\begin{figure}[tbp]
\centering 
\includegraphics[width=\textwidth]{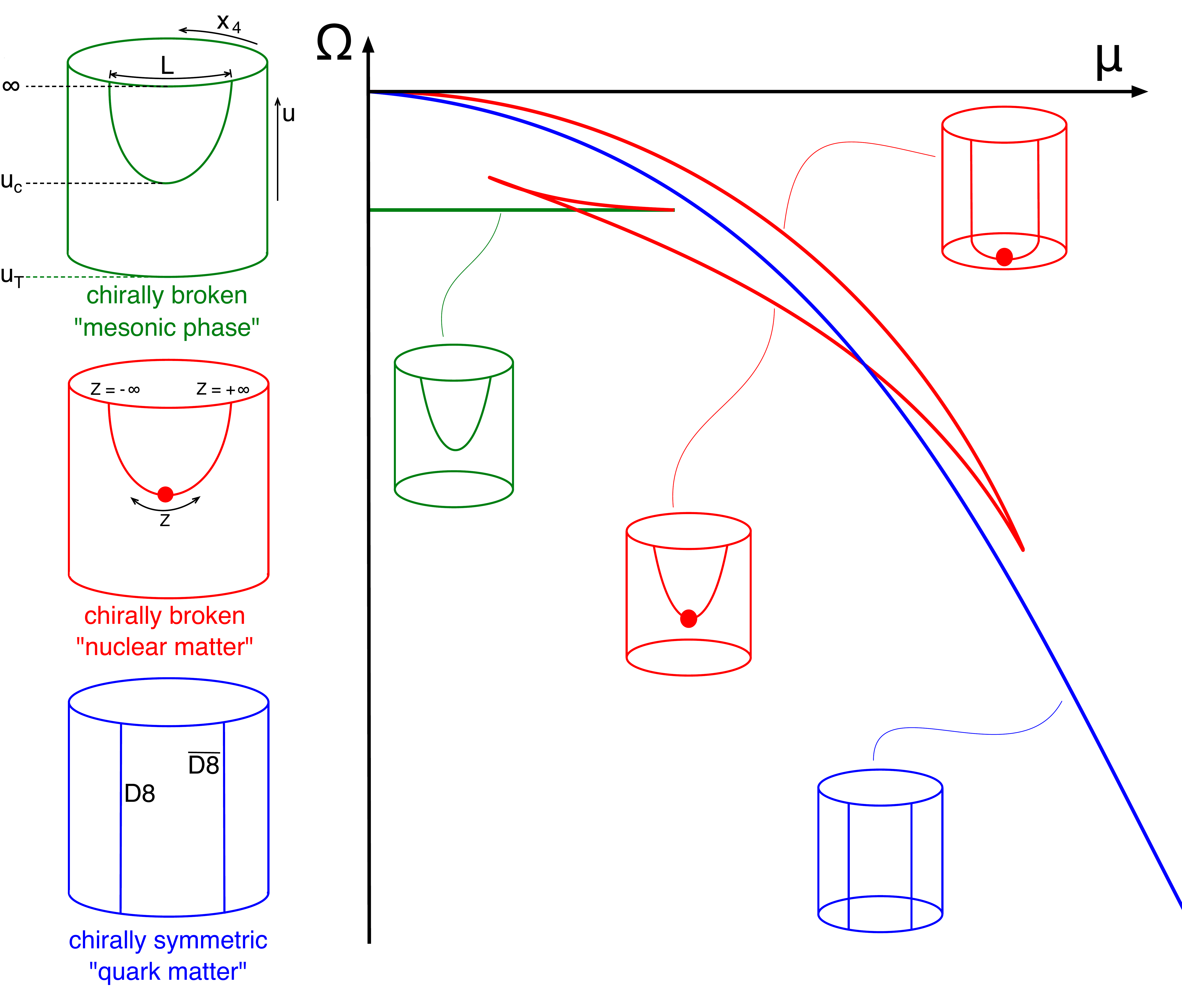}
\caption{\label{fig:schematic} Schematic plot of the free energy $\Omega$ as a function of the chemical potential $\mu$, 
showing the continuous path that connects nuclear matter with quark matter. The various branches are labelled by the geometry of the flavor branes in the $(u,x_4)$ subspace, with the dot on the connected flavor branes indicating the instantons. 
} 
\end{figure}

\begin{figure}[tbp]
\centering
$z$ vs. $x_1$\hspace{2.1cm} $x_2$ vs. $x_1$\hspace{2.2cm} $u$ vs. $x_4$\hspace{2.3cm}$d$ vs. $\mu$  \hspace{0.2cm}

\includegraphics[width=\textwidth]{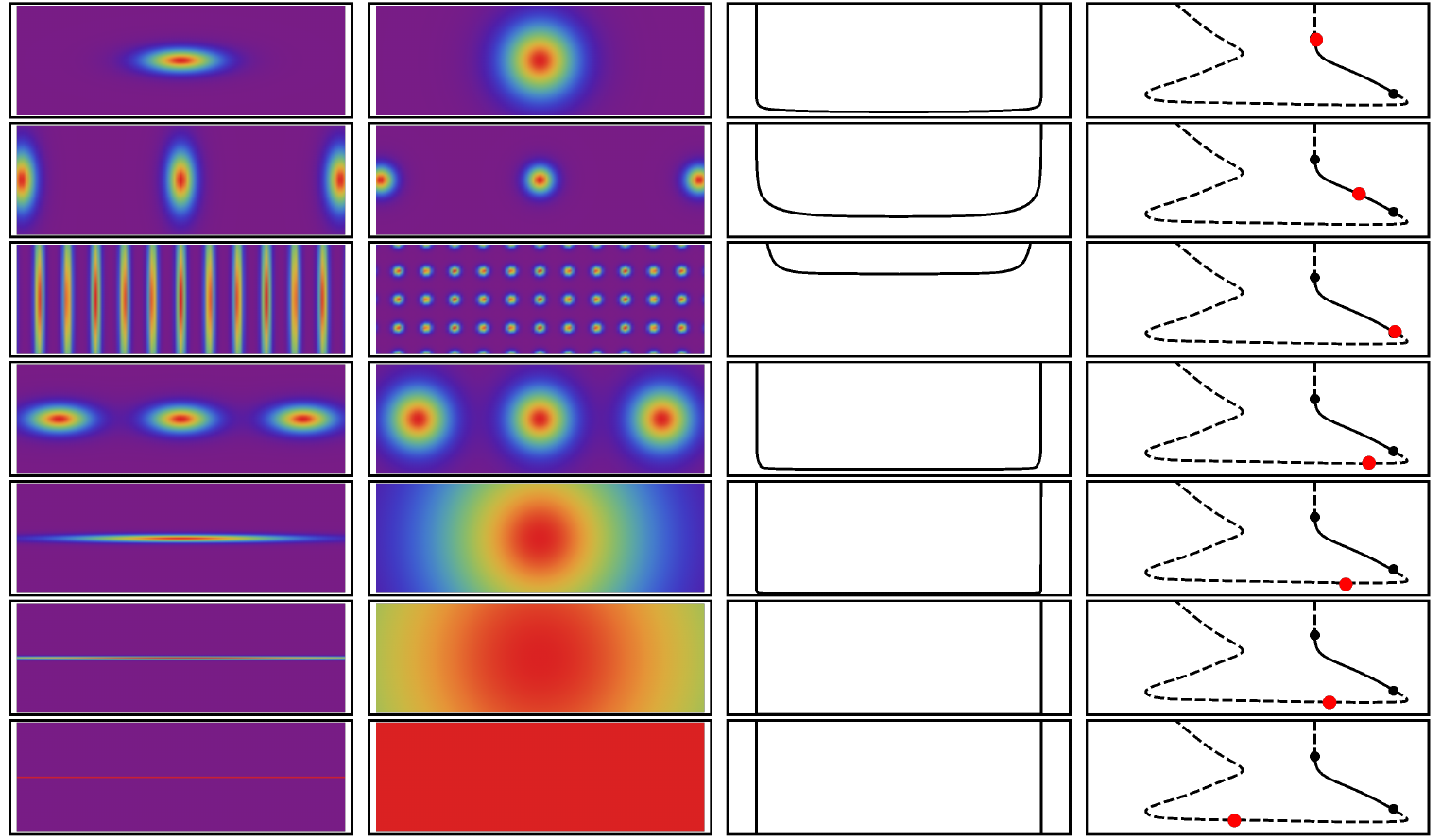}
\caption{\label{fig:series} {\it First column:} instanton profiles cut along the $z$-$x_1$ plane at $x_2=x_3=0$, with $z\in[-2,2]$ and an $x_1$ range of length 2. {\it Second column:} cubic instanton lattice in the $x_1$-$x_2$ plane at $z=x_3=0$, with an $x_1$ range of length 
2 and $x_2$ range of length 2/3. {\it Third column:} embedding function of the flavor branes, $x_4\in[-0.6,0.6]$ and $u\in [0,10]$, showing the continuous transition from the chirally broken to the chirally restored phase. The branes are asymptotically fixed at $x_4(u=\infty)=\pm 1/2$. {\it Fourth column:} same double-logarithmic plot as in the left panel of Fig.\ \ref{fig:dP}, with the large (red) dot indicating the $(d,\mu)$ values for each row and the line style distinguishing between stable (solid) and metastable or unstable (dashed) phases.}
\end{figure}

Our results show a continuity between nuclear and quark matter in the following sense: the curve $[k(\mu),u_c(\mu),n_I(\mu)]$ 
describing the stationary points of the free energy is continuous and is a path that 
can be traced from the baryonic phase into the quark matter phase (along which $\mu$ is non-monotonic). This continuous 
path manifests itself also in the free energy, which is shown schematically in Fig.\ \ref{fig:schematic}. We have chosen 
a schematic representation for convenience, which is a distorted, but not disrupted, version of the path obtained from the 
actual calculation. We have also indicated the geometries of the flavor branes in the two-dimensional $(u,x_4)$ subspace of the model, together with a "legend" that explains these geometries. If we follow the state with the lowest free energy, the transition from nuclear to quark matter is {\it not} continuous. For instance $n_I$, which is the (negative) derivative of  
$\Omega$ with respect to $\mu$, as well as $u_c$ are discontinuous. Therefore, the continuous path just described passes through metastable and unstable states. At zero chemical potential, the baryonic phase connects continuously to the chirally symmetric phase. With the help of the analytical solutions 
(\ref{smallMu}) this statement is exact and does not rely on a numerical calculation: the small-$\mu$ expansions show for instance that $u_c$ indeed vanishes as $\mu$ goes to zero. We can investigate this small-$\mu$ regime 
in more detail for instance by studying the behavior of the instantons. In Fig.\ \ref{fig:series} we show the 
instanton profiles at certain locations of the continuous path.

This figure shows the instanton profiles via cuts in the $(z,x_1)$ and $(x_1,x_2)$ subspaces in the first two columns. 
The profiles are given by the field strengths squared from the first line of Eq.\ (\ref{Fsum}), and in each row the color 
scale is adjusted to the maximal value of the profile. For simplicity, we have 
calculated the profiles from the non-interacting limit, which makes almost no difference since, as we have seen and as this figure further illustrates, the instantons never overlap significantly. 
The figure 
also shows the corresponding embedding of the flavor branes (third column) and indicates the value of $\mu$ and $d$ 
for each row (fourth column). The first three rows correspond to stable solutions, the first row being located at the baryon onset and the third row at the chiral phase transition. As we move towards the chirally symmetric phase the instantons become infinitesimally thin in the holographic direction, while they spread out in the spatial direction and become infinitely wide\footnote{From a top-down string-theoretical point of view one might argue that the resulting large derivatives of the field strengths require derivative corrections of higher order in $\alpha'$ to the DBI action. Here we do not include such terms for simplicity.}. This is a consequence of the scaling of the holographic width $\rho\propto u_c^{3/4}$ and the spatial width $\rho/\gamma \propto u_c^{-3/4}$. Since $u_c$ goes to zero according to Eq.\ (\ref{smallMu}) we have $\rho\propto \mu^{3/2}$ and  $\rho/\gamma \propto \mu^{-3/2}$.  Note that nonzero-temperature effects will slightly change this picture. The geometry dictates that 
$u_c\ge u_T$ and thus $u_c>0$ for $T>0$. Hence the instantons will always retain a nonzero width in the holographic direction and a finite spatial width within the present ansatz.
We leave a more thorough study of nonzero temperatures to the future.

Having emphasized the continuous connection between nuclear and quark matter, let us discuss and interpret this result. It is striking that the connection relies on the nuclear and quark matter branches to "meet" at zero chemical potential. The reason why the quark matter solution exists all the way down to $\mu=0$ is that we work in the chiral limit and thus quarks can be placed into the system at infinitesimally small chemical potential. Of course, at small chemical potential 
the quark matter phase is energetically disfavored compared to the mesonic phase (in real-world QCD one would expect an infinite energy cost due to confinement). Nucleons, on the other hand, do have a mass and are placed into the system at a nonzero chemical potential. Therefore, the uppermost branch in Fig.\ \ref{fig:schematic} should not be confused with ordinary nuclear matter where the density is reduced as we approach the origin. 
Ordinary low-density nuclear matter sits on the lower branch and is superseded by the mesonic phase as the chemical potential is decreased. We thus emphasize that we are not simply decreasing the density of two phases until all matter is gone and then claim a continuous connection between them. We rather observe that all branches, from the mesonic phase through the nuclear matter phase up to the quark matter phase are continuously connected if we include all solutions of the equations of motion and all stationary points of the free energy, which are not necessarily stable. In fact, the uppermost branch in Fig.\ \ref{fig:schematic} can be expected to be unstable
because it has the largest free energy of all solutions we have found. It should thus 
be a local maximum, not a local minimum of the free energy. By plotting the free energy for a fixed $\mu$ as a function of $n_I$ and $u_c$, with $k$ determined such that the constraint (\ref{dk}) is fulfilled, 
we have confirmed this expectation. 
Despite being unstable, the existence of this branch is interesting for the following reason. Ultimately, we are interested in a single continuous potential, defined in the entire "order parameter space". If there is a first-order
phase transition between nuclear and quark matter, one should be able to use this potential to connect the two phases continuously.
The knowledge of the full potential is for instance needed for the calculation of the surface tension of an interface between nuclear and quark matter \cite{Palhares:2010be,Fraga:2018cvr}. Our approximation does not yield such a potential. It is easy 
to check that there are regions in parameter space where the potential becomes complex. However, compared to previous approximations within the same model (and compared to the vast majority of field-theoretical models) we have made a step forward by at least finding a path in parameter space along stationary points of the potential that connects nuclear and quark matter. 

Since we are working with massless quarks, the transition from the chirally broken to the chirally restored branch at zero chemical potential can be expected to involve a discontinuity in second derivatives of the free energy, just like an ordinary second-order phase transition. We shall see in the next section that this is indeed the case, by computing the speed of sound. In other words, since chiral symmetry is an exact symmetry, instantons are topologically protected and a transition from a state with nonzero instanton number to a state without instantons must involve some kind of discontinuity. These arguments suggest an obvious improvement for the future, namely to include nonzero quark masses, which breaks chiral symmetry explicitly. In this case, firstly, the quark and nuclear branches may connect without reaching back to zero chemical potential (since 
massive quarks require a finite energy and thus the quark matter branch cannot start at zero chemical potential). And, secondly, the transition between the chirally broken and (approximately) symmetric phases is allowed to be smooth, including the second derivatives of the free energy. It is then even conceivable that the multivalued curve in Fig.\ \ref{fig:schematic} "straightens out" at large densities, such that an actual quark-hadron continuity occurs in the phase diagram. Therefore, and despite the various approximations we have made and despite the various differences to real-world QCD, we believe that our approach can be very useful in understanding the 
transition between nuclear and quark matter from a physical point of view.

\subsection{Speed of sound}
\label{sec:sound}

The general form of the 
speed of sound is 
\bea \label{sound}
c_s^2 &=&  \frac{\partial P}{\partial \epsilon} =\frac{n^2\frac{\partial s}{\partial T}+s^2\frac{\partial n}{\partial\mu}-ns\left(\frac{\partial n}{\partial T}+\frac{\partial s}{\partial \mu}\right)}{(\mu n + sT)\left(\frac{\partial n}{\partial \mu}\frac{\partial s}{\partial T}-  
\frac{\partial n}{\partial T}\frac{\partial s}{\partial \mu}\right)} \, , 
\eea
where the derivative of the pressure $P$ with respect to the energy density $\epsilon=-P+\mu n + sT$ is taken at fixed entropy per particle $s/n$, and 
where the derivatives of entropy density $s$ and baryon number density $n$ with respect to the baryon chemical potential $\mu$ are taken at fixed temperature $T$ and vice versa. We derive 
Eq.\ (\ref{sound}) in appendix \ref{app:soundgeneral}, see also Refs.\ \cite{Herzog:2008he,Alford:2012vn} for similar derivations in the context of superfluids and appendix A of Ref.\ \cite{Floerchinger:2015efa}. 
As above, we restrict ourselves to zero temperature. In this case
the speed of sound in the baryonic phase and the chirally restored phase becomes  
\be\label{nmu}
c_s^2 = \frac{n}{\mu}\left(\frac{\partial n}{\partial \mu}\right)^{-1} \, ,
\ee
which is a useful relation to keep in mind for the following results. 
Since a phase can only be thermodynamically stable if its density increases monotonically with the corresponding chemical potential, this relation also shows that a negative speed of sound squared indicates an instability (assuming that $n$ and $\mu$ themselves are both positive). 
We compute the speed of sound numerically for the 
baryonic solution discussed in the previous subsection, i.e., with the parameters from Eq.\ (\ref{parachoice}). 
The results for the mesonic phase and the quark matter phase are derived in appendix \ref{app:sound} and read 
\begin{subequations}
\bea
\mbox{mesonic:} \qquad c_s^2(\mu,T\to 0) &=& \frac{1}{5} \, , \\[2ex]
\mbox{chirally symmetric:} \qquad c_s^2(\mu,T) &=& \frac{2}{5}\frac{u_T\sqrt{n_I^2+u_T^5}(n_I^2+5u_T^5)+\mu n_I(n_I^2+6u_T^5)}{(n_I^2+6u_T^5)(\mu n_I+2u_T\sqrt{n_I^2+u_T^5})} \non[2ex]
&=& \left\{\begin{array}{cc} \displaystyle{\frac{1}{6}} & \;\;\mbox{for}\; \mu=0
\\[2ex] \displaystyle{\frac{2}{5}} & \;\;\mbox{for}\; T=0 \end{array}\right. \, . \label{cs2}
\eea
\end{subequations}
The result for the mesonic phase away from the zero-temperature limit can only be computed numerically, but is not needed here. The general result for the chirally symmetric phase is written in terms of the baryon density $n_I$, which depends on $\mu$ and $T$ in a non-analytical way. (Although there are no instantons in the chirally symmetric phase we have kept denoting the dimensionless baryon density by $n_I$ for consistency.)

\begin{figure}[tbp]
\centering
\includegraphics[width=0.7\textwidth]{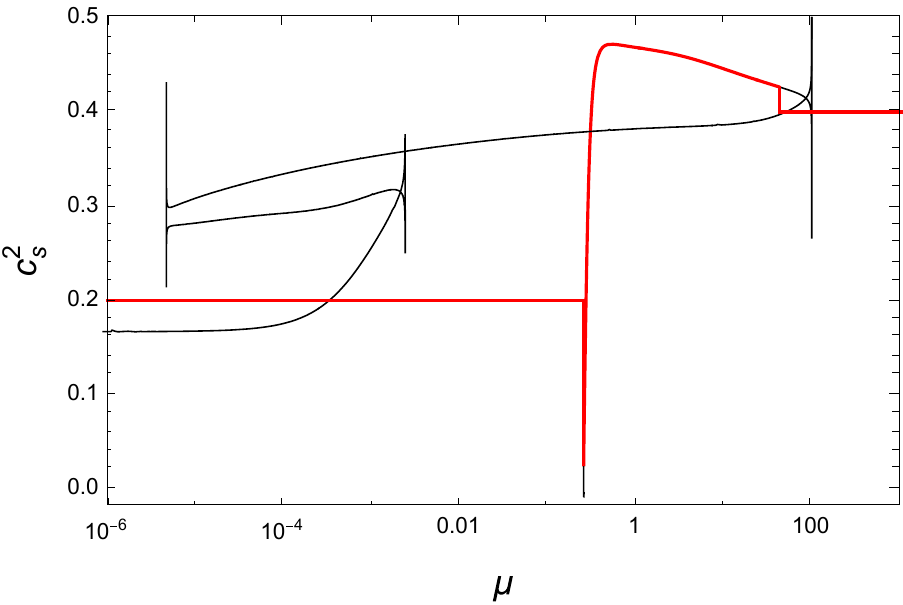}
\caption{\label{fig:sound} Speed of sound squared $c_s^2$ in units of the speed of light as a function of the dimensionless quark chemical potential $\mu$, computed with the parameters (\ref{parachoice}). The thick (red) curve shows the result for the stable phase, from the mesonic phase at low $\mu$ through the nuclear matter phase at intermediate $\mu$ to the quark matter phase at large $\mu$. The thin (black) curve corresponds to metastable and unstable phases. }
\end{figure}

In Fig.\ \ref{fig:sound} we show the speed of sound squared 
as a function of chemical potential in the baryonic phase and, in the regimes where they are favored, in
the mesonic and quark matter phases.  Close to the baryon onset, barely visible on the plot, $c_s^2$ is negative. The solution is unstable for densities smaller than about 66\% of the saturation density, then becomes metastable and finally stable at saturation. 
Besides this instance, there are three other turning points of the curve $n_I(\mu)$ close to which its derivative is negative. In contrast to the turning point close to the baryon onset, in these three instances the curve $n_I(\mu)$ goes through a point where its derivative is zero such that, according to Eq.\ (\ref{nmu}), the speed of sound diverges. Apart from small vicinities of the turning points the speed of sound is real for the entire 
nuclear matter solution. We should keep in mind, however, that $c_s^2>0$ is a necessary, but not sufficient, 
stability criterion. As we know from the previous subsection, certain segments of the thin (black) curve in 
Fig.\ \ref{fig:sound} correspond to maxima of the free energy and thus are unstable. 

At very small $\mu$, i.e., in the regime of the continuous transition from the connected to the disconnected flavor branes, 
the speed of sound squared approaches 1/6. This can be seen from the analytical result (\ref{smallMu}), which shows 
$n_I\propto \mu^6$, from which 
$c_s^2=1/6$ follows with the help of Eq.\ (\ref{nmu}). This value is different from the speed of sound in the 
chirally restored phase, which is $c_s^2=2/5$, as shown for large chemical potentials in the figure. Consequently, as we go from the chirally broken to the chirally restored phase at zero chemical potential, the speed of sound is discontinuous. As explained above, it is expected that second derivatives of the free energy show discontinuities due to the 
different symmetries of the phases. Since the speed of sound (\ref{nmu}) contains a second derivative, its discontinuity at zero chemical potential is no surprise. 
Interestingly, 1/6 is the value of the speed of sound in the chirally restored phase if we set $\mu=0$ and then take the limit $T\to0$ (note the different values of $c_s$ at $T=\mu=0$ depending on the order in which the limits of the function in Eq.\ (\ref{cs2}) are taken). 

Let us now comment on the stable (red) branch. At the baryon onset, the speed of sound jumps from its mesonic value 
down to $c_s^2\simeq 0.025$, then increases to a maximum of about $c_s^2\simeq 0.47$, then decreases and finally jumps down to 
the constant value of the quark matter phase at the chiral transition. In QCD, we know that for asymptotically large $\mu$ the speed of sound assumes the conformal value $c_s^2=1/3$ and perturbative corrections decrease this value. We also know 
that nuclear matter at saturation has a much smaller -- non-relativistic -- speed of sound, close to the value we have found here. It is unknown how the low-density and high-density regimes are connected. It has been pointed out that the simplest scenario, 
connecting the two regimes with a monotonic function, creates tension with astrophysical observations \cite{Bedaque:2014sqa}. More precisely, a sufficiently stiff equation of state (= a sufficiently large speed of sound) is required to obtain observed neutron star masses of twice the solar mass \cite{Demorest:2010bx,Antoniadis:2013pzd}, possibly larger
\cite{Linares:2018ppq}. Hence, 
in the intermediate density regime the speed of sound most likely has to exceed the conformal value. 
There have been suggestions in the literature that the conformal limit may be an absolute upper bound, but counterexamples have been pointed out within the gauge/string duality \cite{Hoyos:2016cob,Ecker:2017fyh}. Values larger than $c_s^2>1/3$ are routinely reached in 
phenomenological models of nuclear matter or in extrapolations of low-density effective theories, but in these calculations the speed of sound typically behaves monotonically and cannot be traced into the chirally symmetric phase. Therefore, phenomenologically motivated interpolations between the low-density and high-density QCD results have been studied and the consequences for masses and radii of neutron stars have been tested \cite{Tews:2018kmu}. It is intriguing that the nuclear matter of our holographic calculation does show a non-monotonic behavior, as required from putting together constraints from QCD and neutron stars. Of course, our result is different from QCD at asymptotically large $\mu$. The reason is that 
the mass scale $M_{\rm KK}$ plays a role even in this asymptotic regime, allowing 
the speed of sound to be different from $1/3$. This might be interpreted as a non-perturbative strong-coupling effect, having in mind that we need to stop trusting the model for asymptotically large densities, where QCD exhibits asymptotic freedom. We also mention again that the chiral transition occurs at an extremely large density, a fact easy to be overlooked due to the use of the logarithmic scale in Fig.\ \ref{fig:sound}, and, as explained above, that one should treat the quantitative interpretation of the model with some care, especially at large densities. Nevertheless, our approach allows for a consistent microscopic calculation of nuclear and quark matter with a qualitative prediction for the speed of sound that is very interesting in view of recent astrophysical constraints.

\section{Summary and outlook}
\label{sec:conclusions}

We have shown that chirally broken and chirally symmetric phases in the holographic Sakai-Sugimoto model can be continuously connected in the presence of instantons on the flavor branes. Geometrically, this continuity is realized by transforming the curved, connected flavor branes for left- and right-handed fermions into straight, disconnected branes. The transformation appears dynamically by solving 
the equations of motion and computing the embedding of the flavor branes in the deconfined geometry of the model and in the presence of an interacting many-instanton 
system. We have approximated the instanton interactions with the help of the exact flat-space two-instanton solution.
We have found
that, as the connected flavor branes become straight and approach the disconnected embedding, the instantons become 
infinitesimally thin in the holographic direction, but spread out to become infinitely wide in the spatial direction. 
Nevertheless, they barely overlap since in this limit also their density goes to zero. Most previous related studies had
included instantons only in the confined geometry or did not include interactions between the instantons, i.e.,
to the best of our knowledge this continuity has been observed for the first time in the present paper. 

Our observation is closely related to the transition between nuclear matter (= instanton system with connected flavor branes) 
and chirally symmetric quark matter (= disconnected flavor branes) in dense QCD. While our continuity does not directly connect stable nuclear matter at the lowest densities with stable ultra-dense quark matter, it does so on a path that goes through metastable and unstable  
stationary points of the potential. In particular, the actual phase transition between nuclear and quark matter turns out to be of first order. Since we have worked in the chiral limit, it is obvious that the actual transition cannot be continuous, since chiral symmetry is exact. It would thus be very interesting to include quark masses into our approach and 
see whether and how the continuity is affected. This can be done with worldsheet instantons \cite{Hashimoto:2008sr}, which have been used to study the effect of quark masses on the hadron spectrum \cite{Aharony:2008an,Hashimoto:2009hj,Hashimoto:2009st,Bigazzi:2018cpg}, or with a tachyonic effective action \cite{Bergman:2007pm,Dhar:2007bz}.

Besides this main theoretical result we have also pointed out phenomenological applications of our approach. We have shown that it is possible to fit the parameters of the model to reproduce the basic properties of isospin-symmetric nuclear matter at saturation. In doing so, it was crucial to introduce two additional parameters that characterize the instanton shape, increasing the 
number of model parameters from 3 to 5. We have worked with a single parameter set, and future studies are required for a more 
systematic investigation of the parameter space. For instance, we have seen that for the chosen parameters the chiral transition occurs at an extremely large baryon density -- probably unrealistically large, at least compared to predictions of other models.
It would be interesting to see whether different parameter sets can be found that are also in agreement with nuclear physics and yield different 
values for the chiral transition. We have also calculated the speed of sound to further connect our 
results to the phenomenology of dense matter. Our result shows a non-monotonic behavior of the speed of sound in nuclear matter that 
reaches a maximum larger than that of the quark matter phase, before it jumps to the quark matter value at the chiral transition. Interestingly, this non-monotonic behavior is suggested by putting together constraints from 
QCD and astrophysical data, most notably the largest observed neutron star mass of about two solar masses. It is therefore a natural extension of our work to compute the equation of state and the 
resulting maximum mass of a star that can be reached with it. 

Let us conclude with mentioning some further extensions that may improve our calculation. Obvious technical improvements concern our 
many-instanton approximation. We have constructed the interaction terms solely from 2-instanton contributions, and those, in turn, are based on the flat-space solutions. Moreover, we have simply averaged over position space before solving the equations of motion and the embedding of the flavor branes. Also, we have kept the background geometry fixed although the gauge fields 
on the flavor branes become very large at large densities and backreactions may be non-negligible. Improvements on all these points are difficult, but since we are working within a top-down approach it is in principle known how to proceed. More direct  extensions are for instance the study of two- or multi-layer solutions, which we have briefly discussed but not included into our numerical evaluation; isospin-breaking terms, which are needed for a more realistic study of neutron star matter; 
and nonzero-temperature effects, which we have included in almost all our equations, but not in the 
numerical results.

\acknowledgments

We would like to thank N.\ Evans, A.\ Rebhan, and S.\ Reddy for valuable comments.
A.S.\ is supported by the Science \& Technology Facilities Council (STFC) in the form of an Ernest Rutherford Fellowship. KBF thanks the University of Southampton for their hospitality during his sabbatical.

\appendix

\section{Instantons from the ADHM construction}
\label{app:flat}

In this appendix we derive the two-instanton solution from the ADHM construction \cite{Atiyah:1978ri}. 
This is in large parts a recapitulation of derivations from Refs.\ \cite{Kim:2008iy, Hashimoto:2009ys}, adapted to our purposes. In particular, we derive Eq.\ (\ref{calFsq}) in the main part. 

According to the ADHM construction, the gauge fields for an Sp($n$) $k$-instanton solution, 
where Sp($n$) is the unitary symplectic group, are
\be \label{AM}
A_M = -iU^\dag\partial_M U \, , 
\ee
where $M=1,2,3,z$, and $U$ is a $(n+k)\times n$ matrix with quaternionic entries, which satisfies 
\be \label{unit}
U^\dag U = {\bf 1}_{n} \, ,
\ee
and
\be \label{DU0}
\Delta^\dag U=0 \, .
\ee
Here, $\Delta$ is a quaternionic $(n+k)\times k$ matrix, such that the quaternionic $k\times k$ matrix
\be
L\equiv \Delta^\dag \Delta
\ee
is symmetric and real, i.e., it commutes with quaternions. 
The relevant case for us will be $n=1$ since Sp(1) $\cong$ SU(2). 
We write the quaternions as 
\be
q=q_Me_M \, , 
\ee
with $M=1,2,3,4$, $q_M\in \mathbb{R}$, and the basis vectors $e_i = i\sigma_i$ for $i=1,2,3$ and $e_4 = 1$, where $\sigma_i$ are the Pauli matrices, and the 4-direction corresponds to the holographic $z$ coordinate. We denote $q^\dag = q_M\bar{e}_M$ with $\bar{e}_i=-i\sigma_i$ for $i=1,2,3$ and $\bar{e}_4=1$. 
Below we shall need the 
scalar product and cross product of two quaternions, defined as  
\begin{subequations} \label{scalarcross}
\bea
q\cdot p &\equiv& \frac{1}{2}(q^\dag p +p^\dag q) =  q_Mp_M \, , \\[2ex]
q\times p &\equiv& \frac{1}{2}(q^\dag p -p^\dag q) =  i(q^4\vec{p}-p^4\vec{q}+\vec{q}\times\vec{p}
)\cdot\vec{\sigma} \, ,
\eea
\end{subequations}
where we have used $\{\sigma_i,\sigma_j\}=2\delta_{ij}$, $[\sigma_i,\sigma_j]=2i\epsilon_{ijk}\sigma_k$.
Note in particular that the scalar product is proportional to the unit matrix, while the cross product is spanned by the other three basis vectors of the quaternionic space. It is important to keep in mind that  the $2\times 2$ space introduced by the representation of the quaternions occurs for 
every gauge group, it has nothing to do with the SU(2) gauge group relevant for our case. We will represent the 
four-dimensional vector $x_M=(\vec{x},z)$ as a quaternion as $x = x_Me_M = i\vec{x}\cdot\vec{\sigma}+z$, 
such that $e_M=\partial_Mx$.

The matrix $\Delta$ can be written as $\Delta = a+b\otimes x$ with quaternionic $(n+k)\times k$ matrices $a$, $b$ that are constant in $x$, and where in $b\otimes x$ each element of $b$ is multiplied by $x$ from the right. One can choose $b^\dag =(0_{k\times n},-{\bf 1}_{k})$, while $a$ remains generic and encodes the widths and locations of the instantons (more precisely, $a$ depends on parameters that correspond to widths and locations of the instantons
if the instantons are sufficiently far away from each other). 
Thus, 
with a quaternionic $k\times k$ matrix $X$ and a quaternionic $n\times k$ matrix $Y$ we can write  
\be \label{DL}
\Delta = \left(\begin{array}{c} Y \\ -(x-X) \end{array}\right) \quad \Rightarrow \qquad L=Y^\dag Y+(x-X)^\dag(x-X) \, .
\ee
Using Eq.\ (\ref{AM}) and $F_{MN} = \partial_M A_N -\partial_N A_M +i[A_M,A_N]$ the field strengths 
become 
\bea \label{FMN}
F_{MN}&=&-i\partial_M U^\dag(1-UU^\dag)\partial U_N+i\partial_N U^\dag(1-UU^\dag)\partial U_M \non[2ex]
&=& -iU^\dag(\partial_M\Delta)L^{-1}(\partial_N\Delta^\dag)U+iU^\dag(\partial_N\Delta)L^{-1}(\partial_M\Delta^\dag)U \non[2ex]
&=&-2iU^\dag b[L^{-1}\otimes (\bar{e}_M\times \bar{e}_N)] b^\dag U \, .
\eea
Here we have used Eq.\ (\ref{unit}), the identity $UU^\dag = 1-\Delta L^{-1} \Delta^\dag$ \cite{Corrigan:1979di,Corrigan:1983sv}, Eq.\ (\ref{DU0}),  and the derivatives $\partial_M\Delta = b\otimes e_M$, $\partial_M\Delta^\dag=\bar{e}_M\otimes b^\dag$, where in $\bar{e}_M\otimes b^\dag$ every element of $b^\dag$ is multiplied from the left by $\bar{e}_M$. Moreover, we have used that $L^{-1}$ commutes with quaternions. The result (\ref{FMN}) can be written with the
help of the 't Hooft symbol $\eta$,
\be
\bar{e}_M\times \bar{e}_N = \frac{1}{2}(e_M\bar{e}_N-e_N\bar{e}_M) = i\eta_{iMN}\sigma_i \, , 
\ee
where $\eta_{ijk}=\epsilon_{ijk}$, $\eta_{ij4}=-\eta_{i4j}=\delta_{ij}$, $\eta_{i44}=0$. Consequently, the ADHM construction ensures the self-duality of the field strengths,  $F_{MN}=\tilde{F}_{MN}=\frac{1}{2}\epsilon_{MNRS}F_{RS}$ (using $\epsilon_{ijk4}=+\epsilon_{ijk}$),
or
\be
F_{iz}=\frac{1}{2}\epsilon_{ijk}F_{jk} \, .
\ee 
A useful relation for the trace over the field strengths is \cite{Osborn:1979bx}
\be \label{Boxsq}
\Tr[F_{MN}^2] 
= -\Box^2\ln{\rm det} \,L \, ,
\ee
where $\Box=\partial_M^2$. Below we shall compute $F_{MN}^2$ explicitly for the 2-instanton solution and thus we do not really need to employ this relation. Nevertheless, calculating the trace of the field strength squared via $L$ can be used as a check for our 
explicit form. 

From now on we will only consider $n=1$, i.e., SU(2) instantons.

\subsection{Single instanton}

As a warm up, it is useful to start with a single instanton. In this case, $k=1$ and thus $X$ and $Y$ are quaternions. 
Anticipating the notation used for $k=2$ below let use denote $y=Y$. We determine $U$ by writing $U^\dag = (\alpha^\dag, \beta^\dag)$ with quaternions 
$\alpha$, $\beta$ (in general, $\alpha$ is a quaternionic $n\times n$ matrix and $\beta$ is a quaternionic $k\times n$ matrix). Then,  Eq.\ (\ref{DU0}) reads
\be
0 = \Delta^\dag U = y^\dag \alpha -(x-X)^\dag \beta \, .
\ee
After multiplying this equation by $y$ from the left we obtain $\alpha$, which we insert into 
Eq.\ (\ref{DU0}) to obtain $|\beta|^2$. Denoting $\hat{\beta}=\beta/|\beta|$ we thus find
\be
U^\dag = \frac{\rho}{\sqrt{\xi^2+\rho^2}}(\hat{\beta}^\dag (x-X)y^{-1}, \hat{\beta}^\dag) \, , 
\ee
where $\xi^2 = |x-X|^2$ and $\rho^2 = |y|^2$, and where we have used $y^\dag=\rho^2 y^{-1}$. 
Inserting this into the gauge field (\ref{AM}), we compute
\bea
A_M &=& -i\hat{\beta}^\dag(fg\partial_M g^{-1})\hat{\beta} \, ,
\eea
where $f\equiv \xi^2/(\xi^2+\rho^2)$, $g\equiv (x-X)/\xi$.
This is the well-known BPST solution.
The field strengths can easily be computed from Eq.\ (\ref{FMN}). In the single-instanton case $L=\rho^2+\xi^2$ is a scalar and taking its inverse is trivial. We find
\be \label{Fizapp}
F_{MN}= 2\eta_{aMN}\frac{\rho^2\hat{\beta}^\dag\sigma_a\hat{\beta}}{(\xi^2+\rho^2)^2} \, .
\ee
With $\eta_{aMN}\eta_{bMN}=4\delta_{ai}\delta_{bi}$ we obtain
\be \label{FMNsq1}
F_{MN}^2 = \frac{16\rho^4\sigma_a\sigma_a}{(\rho^2+\xi^2)^4} \, , 
\ee
which is confirmed by Eq.\ (\ref{Boxsq}),
\be
\Tr[F_{MN}^2] = -\Box^2\ln{\rm det}\, L = -\Box^2\ln (\rho^2+\xi^2) = \frac{96\rho^4}{(\rho^2+\xi^2)^4} \, .
\ee
Eq.\ (\ref{FMNsq1}) is the form of the single instanton solution used in the text, 
see Eq.\ (\ref{Fiz0}), where the coordinate $z$ has been rescaled with $\gamma$ (because of self-duality, $F_{MN}^2 = 4F_{iz}^2$). 

\subsection{Two instantons}

For the two-instanton solution we write  
\be
Y=(y_1,y_2) \, , \qquad X = \left(\begin{array}{cc}X_1 & w \\ w& X_2\end{array}\right) \quad \Rightarrow \qquad  \Delta = \left(\begin{array}{cc} y_1 & y_2 \\ -(x-X_1) & w \\ w & -(x-X_2) \end{array}\right) \, ,
\ee
with quaternions $y_1$, $y_2$, $X_1$, $X_2$, $w$, from which we compute $L$ according to Eq.\ (\ref{DL}),
\bea \label{Ltwo}
L &=&  \left(\begin{array}{cc} |y_1|^2 +|x-X_1|^2+|w|^2 & y_1^\dag y_2-(x-X_1)^\dag w-w^\dag(x-X_2) \\[2ex] y_2^\dag y_1-w^\dag(x-X_1)-(x-X_2)^\dag w &
|y_2|^2 +|x-X_2|^2+|w|^2 \end{array}\right) \non [2ex]
&=& \left(\begin{array}{cc} \rho_1^2 +|x-X_1|^2+\frac{\rho_1^2\rho_2^2-(y_1\cdot y_2)^2}{|X_1-X_2|^2} & -2x\cdot w +y_1\cdot y_2 +\frac{2(X_1^\dag X_2)\cdot (y_2\times y_1)}{|X_1-X_2|^2}  \\[2ex] -2x\cdot w +y_1\cdot y_2 +\frac{2(X_1^\dag X_2)\cdot (y_2\times y_1)}{|X_1-X_2|^2} &
\rho_2^2 +|x-X_2|^2+\frac{\rho_1^2\rho_2^2-(y_1\cdot y_2)^2}{|X_1-X_2|^2} \end{array}\right) \, .
\eea
The second line is obtained as follows. From the requirement that $L$ be symmetric, we find 
\be \label{w}
w = \frac{X_1-X_2}{|X_1-X_2|^2} (y_2\times y_1) +c(X_1-X_2) \, , 
\ee
where we set $c=0$ \cite{Hashimoto:2009ys}. This yields 
\be
|w|^2 = \frac{\rho_1^2\rho_2^2-(y_1\cdot y_2)^2}{|X_1-X_2|^2} \, , 
\ee
where we have defined $\rho_1\equiv|y_1|$, $\rho_2\equiv|y_2|$.  Finally, we have rewritten
\be
y_1^\dag y_2+X_1^\dag w+w^\dag X_2  = y_1\cdot y_2 +\frac{2(X_1^\dag X_2)\cdot (y_2\times y_1)}{|X_1-X_2|^2} \, ,
\ee
where the right-hand side is manifestly real because it is composed of scalar products of quaternions. 

We can now compute the field strengths analogously to the case of the single instanton. We first need to compute $U^\dag=(\alpha^\dag,\beta^\dag)$. For the two-instanton solution, $\alpha$ remains a quaternion, while 
$\beta$ becomes a $2\times 1$ quaternionic matrix, which we write as $\beta^\dag = (\beta_1^\dag, \beta_2^\dag)$ with quaternions $\beta_1$, $\beta_2$. Then we can express Eq.\ (\ref{DU0}) solely through quaternions, 
\be
0 = \Delta^\dag U = \left( \begin{array}{ccc}y_1^\dag & -(x-X_1)^\dag & w^\dag \\ y_2^\dag & w^\dag & -(x-X_2)^\dag \end{array}\right) \left(\begin{array}{c} \alpha \\ \beta_1 \\ \beta_2 
\end{array}\right) \, .
\ee 
We manipulate the resulting two quaternionic equations as follows. We multiply the first one from the left with $y_1$, and the second one with $y_2$, and we define $\hat{y}_i=y_i/\rho_i$ for $i=1,2$. Then, the first equation yields
\be \label{alpha}
\alpha = \frac{\hat{y}_1}{\rho_1}[(x-X_1)^\dag \beta_1 - w^\dag \beta_2] \, , 
\ee
and subtracting the second from the first equation gives
\be \label{Pbeta}
P_1 \beta_1 = P_2\beta_2 \, , 
\ee
where we have defined
\be
P_{1/2} \equiv \frac{\hat{y}_{1/2}}{\rho_{1/2}}(x-X_{1/2})^\dag + \frac{\hat{y}_{2/1}}{\rho_{2/1}}w^\dag \, .
\ee
We multiply Eq.\ (\ref{Pbeta}) from the left with $P_1^\dag$,
and insert the result into Eq.\ (\ref{alpha}) to obtain 
\be
\alpha = \frac{Q}{|P_1|^2} \beta_2 \, , 
\ee
with the definition 
\be
Q\equiv \frac{\hat{y}_1}{\rho_1}\left[(x-X_1)^\dag P_1^\dag P_2-|P_1|^2w^\dag\right] \, .
\ee 
From the condition $1=U^\dag U = |\alpha|^2 + |\beta_1|^2+|\beta_2|^2$ we find
\be
|\beta_2|^2 = \frac{|P_1|^4}{|Q|^2+|P_1|^2|P_2|^2+|P_1|^4} \, , 
\ee
and thus we obtain the final result for $U$, 
\be
U = \frac{1}{\sqrt{|Q|^2+|P_1|^2|P_2|^2+|P_1|^4}}\left(\begin{array}{c} Q \\ P_1^\dag P_2 \\ |P_1|^2 
\end{array}\right)\hat{\beta} \, ,
\ee
where we have denoted $\hat{\beta}\equiv \hat{\beta}_2$ for brevity. We can insert this result into Eq.\ (\ref{FMN}) to compute the field strengths. Denoting the elements of $L^{-1}$ by 
$L^{-1}_{ij} \in \mathbb{R}$, with $L_{12}^{-1}=L_{21}^{-1}$, we obtain
\be \label{FPP}
F_{MN} = 2\eta_{aMN}\hat{\beta}^\dag\frac{L_{11}^{-1}P_2^\dag P_1\sigma_a P_1^\dag P_2 +L_{12}^{-1}|P_1|^2(P_2^\dag P_1\sigma_a+\sigma_a P_1^\dag P_2) +L_{22}^{-1}|P_1|^4\sigma_a}{|Q|^2+|P_1|^2|P_2|^2+|P_1|^4} \,\hat{\beta}\, .
\ee
After some algebra we find
\bea \label{FPP2}
&&F_{MN}^2 = \frac{16|P_1|^4}{(|Q|^2+|P_1|^2|P_2|^2+|P_1|^4)^2} \non[2ex]
&&\times \Big( \sigma_a\sigma_a\Big[(L_{11}^{-1})^2|P_2|^4+(L_{22}^{-1})^2|P_1|^4+4L_{12}^{-1}(L_{11}^{-1}|P_2|^2+L_{22}^{-1}|P_1|^2)P_1\cdot P_2\Big] \non[2ex]
&&+2L_{11}^{-1}L_{22}^{-1}(i\sigma_aP_1^\dag P_2)\cdot(P_1^\dag P_2i\sigma_a)+4(L_{12}^{-1})^2\Big\{|P_1|^2|P_2|^2\sigma_a\sigma_a-[(P_1i\sigma_a)\cdot P_2]^2\Big\}\Big) \, . \hspace{1cm}
\eea
Since all terms are scalar products of quaternions, $F_{MN}^2$ is proportional to the unit matrix
(while $F_{MN}$ isn't), as for the single-instanton case. In contrast to the single instanton, there is a 
nontrivial dependence on the orientations of the instantons $\hat{y}_1,\hat{y}_2 \in {\rm SU(2)}$. 
Since only the relative rotation between the two instantons 
matters, we parameterize \cite{Kim:2008iy}
\be
\hat{y}_1=1 \, , \qquad \hat{y}_2 = e^{i\vec{\theta}\cdot\vec{\sigma}}= \cos\theta +i\hat{\vec{\theta}}\cdot\vec{\sigma}\sin\theta \, ,
\ee
which yields 
\be
\hat{y}_1\cdot \hat{y}_2 = \cos\theta \, , \qquad \hat{y}_1\times \hat{y}_2 = i\hat{\vec{\theta}}\cdot\vec{\sigma}\sin\theta \, .
\ee
An SU(2) transformation $\hat{y}$ can be formulated in terms of a rotation $R\in {\rm SO(3)}$ as follows. Start from an SU(2) matrix $G$ and its transformation $\tilde{G} = \hat{y}^\dag G\hat{y}$, insert 
$G = G_a\sigma_a$ and $\tilde{G}=\tilde{G}_a\sigma_a$, multiply by $\sigma_b$, and take the trace to obtain
\be \label{Rab}
\tilde{G}_b = G_a R_{ab} \, , \qquad  R_{ab} = \frac{1}{2} \Tr[\sigma_a\hat{y}^\dag\sigma_b\hat{y}] \, .
\ee
Thus, in our case, the rotation corresponding to $\hat{y}_2$ is 
\be
R_{ab} = \frac{1}{2} \Tr[\sigma_a  e^{-i\vec{\theta}\cdot\vec{\sigma}} \sigma_b e^{i\vec{\theta}\cdot\vec{\sigma}}] = \delta_{ab}(\cos^2\theta-\sin^2\theta)-2\epsilon_{abc}\hat{\theta}_c\cos\theta\sin\theta
+2\hat{\theta}_a\hat{\theta}_b\sin^2 \theta\, . 
\ee
We now write $X_{1/2}=X\pm D$ and set $X=0$, i.e., the origin of the four-dimensional space is chosen to be the center point of the line that connects the two instantons. In this case, $X_1^\dag X_2=-|D|^2$, and thus $(X_1^\dag X_2)\cdot (y_2\times y_1)=0$ since any cross product is orthogonal to the unit matrix direction, as remarked below (\ref{scalarcross}), and 
from Eq.\ (\ref{Ltwo}) we recover Eq.\ (21) in Ref.\ \cite{Kim:2008iy}.
We choose the separation to be in the $x_1$ direction and to have length $\delta$, i.e., $D=i\delta\sigma_1/2$. This yields
\bea
|x-X_{1/2}|^2=|x|^2\mp \delta x_1 +\frac{\delta^2}{4} \, , \qquad |X_1-X_2|^2 = \delta^2 \, , 
\eea
and
\be
w = \frac{\rho_1\rho_2\sin\theta}{\delta}(\hat{\theta}_1+i\hat{\theta}_2\sigma_3-i\hat{\theta}_3\sigma_2) \, , \qquad |w|^2 = \frac{\rho_1^2\rho_2^2\sin^2\theta}{\delta^2} \, , 
\ee
such that $L$ from Eq.\ (\ref{Ltwo}) is given by the entries
\begin{subequations}
\bea
L_{11/22} &=& \rho_{1/2}^2+|x|^2\mp \delta x_1+\frac{\delta^2}{4}+\frac{\rho_1^2\rho_2^2\sin^2\theta}{\delta^2}  \, , \\[2ex]
L_{12}&=&L_{21}= -2\frac{\rho_1\rho_2\sin\theta}{\delta}(z\hat{\theta}_1+x_3\hat{\theta}_2-x_2\hat{\theta}_3)+\rho_1\rho_2\cos\theta  \, .
\eea
\end{subequations}
We can now invert $L$ and compute the field strength squared (\ref{FPP2}) explicitly. As a check, this can also 
be computed with the help of Eq.\ (\ref{Boxsq}). Here we restrict ourselves to "defensive skyrmions" \cite{Kim:2008iy}, where $\theta=0$, i.e., $R=1$. Also setting $\rho_1=\rho_2\equiv \rho$ for simplicity,
we obtain Eq.\ (\ref{calFsq}) in the main text.

\section{Multiple instanton layers}
\label{app:layers}

If we allow for multiple instanton layers, Eq.\ (\ref{FsqN}) is generalized to 
\bea
F_{iz}^2 &=& \sum_{n=0}^{N_z-1}\sum_{i=1}^{i_n} (F_{iz}^{(1)})^2(\vec{x}_{in},z_n) + \frac{1}{2}\sum_{n=0}^{N_z-1}\sum_{i=1}^{i_n}\sum_{j\neq i}^{i_n} {\cal I}(\vec{x}_{in},z_n,\vec{x}_{jn},z_n) \non[2ex]
&&+\frac{1}{2}\sum_{n=0}^{N_z-1}\sum_{m\neq n}^{N_z-1}\sum_{i=1}^{i_n}\sum_{j=1}^{i_m} {\cal I}(\vec{x}_{in},z_n,\vec{x}_{jm},z_m) \, , 
\eea
where we have written the intra-layer and cross-layer interaction terms separately. We have denoted the number of layers by $N_z$ and their location by $z_n$, and in this general notation each layer is allowed to contain a different number of instantons, denoted by $i_n$.  
From now on we assume all layers to have the same number of instantons, $i_0 = \ldots = i_{N_z-1} \equiv N_{\vec{x}}$, such 
that $N_I=N_{\vec{x}}N_z$ is the total instanton number. Moreover, we neglect the cross-layer interactions. Including them 
does not pose a conceptual problem, we leave this improvement for future work. As in the main part, 
we employ the nearest-neighbor approximation for the intra-layer interactions 
and take the spatial average to obtain 
\be
F_{iz}^2\to  \frac{2\lambda_0^2n_I}{3\gamma N_z}\sum_{n=0}^{N_z-1} \left[(1-p)q_0(z-z_n)+\frac{p}{2}q_{\rm int}(d,z-z_n)\right] \, ,
\ee
which is the generalization of Eq.\ (\ref{Fsum}). Now, since the instanton number in the spatial lattice is different from the 
total instanton number, Eq.\ (\ref{s}) is modified to $v/N_{\vec{x}}= \delta^3/r$, and 
thus Eq.\ (\ref{d}) becomes $d  = \gamma/\rho[6\pi^4 r N_z/(\lambda^2 n_I)]^{1/3}$. Following Ref.\ \cite{Preis:2016fsp}, we may use the following ansatz for the location of the layers,
\be \label{zn}
z_n = \left(\frac{N_z-1}{2}-n\right) \Delta z \, , 
\ee
with $n=0,\ldots , N_z-1$, and where the distance between the layers $\Delta z$ has to be determined dynamically. 
This ansatz is completely general for one and two layers (apart from the obvious assumption that the two layers are symmetric with respect to $z\to -z$). For $N_z>2$ it assumes equidistant layers, which is a simplification in order to 
limit the number of free parameters. It has been shown that within this ansatz and neglecting instanton interactions 
three or more layers are never energetically preferred \cite{Preis:2016fsp}. Although this might be a consequence of the simplistic ansatz, the fact that the same observation was made in a completely different approximation \cite{Elliot-Ripley:2016uwb} suggests that this result might hold more generally.

The Lagrangian (\ref{Lag}) has the same form as in the single-layer case, with the generalized functions
\be
q(u) = 2\frac{\partial z}{\partial u} \frac{1}{N_z}\sum_{n=0}^{N_z-1}\left[(1-p)q_0(z-z_n)+\frac{p}{2}q_{\rm int}(d,z-z_n)\right] \, ,
\ee
and 
\bea 
Q &=& \frac{1}{N_z}\sum_{n=0}^{N_z-1}\left[(1-p)\frac{Q_0(z-z_n)+Q_0(z+z_n)}{2} \right. \non[2ex]
&&+ \left.\frac{p}{2}\frac{Q_{\rm int}(d,z-z_n)+Q_{\rm int}(d,z+z_n)}{2}\right] \, ,
\eea
where we have used that $q_0(z)$ and $q_{\rm int}(d,z)$ are symmetric under $z\to -z$.

Finally, the multi-layer scenario within the ansatz (\ref{zn}) requires an additional equation for the minimization 
of the free energy with respect to $\Delta z$, i.e., we have to add the following equation to the set of equations (\ref{minimOm}),
\bea
0&=&\int_{u_c}^\infty du\, u^{5/2} \left(\frac{g_1\zeta^{-1}+g_2\zeta}{2q}\frac{\partial q}{\partial \Delta z} + \frac{\zeta n_I^2 Q}{u^5}\frac{\partial Q}{\partial \Delta z}\right)
 \, .
\eea
All results in Sec.\ \ref{sec:results} in the main part concern the single-layer case. With the help of the results of this appendix we did check that including multiple layers does not change our main conclusions: the two-layer solution {\it is} 
preferred for large baryon densities, and thus the critical chemical potential for the chiral phase transition is affected, but this does not disrupt the continuity between the chirally broken and chirally symmetric geometries. 

\section{Integrating the interaction term}
\label{app:deriv}

Here we derive the result (\ref{Psi}), i.e., we compute the $z$ integral over the function $q_{\rm int}$ defined in 
Eq.\ (\ref{Phi}). The integration can be done with the help of the following observations.
With the definitions from Eq.\ (\ref{SSab}) we compute 
\begin{subequations}
\bea
\dot{\cal S}_1 &=& \frac{1}{2b\sqrt{a^2+b}}\left[{\cal S}_1(a\dot{b}-\dot{a}b)-\frac{\dot{b}}{2{\cal S}_1}\right] \, , \\[2ex]
\dot{\cal S}_2 &=& \frac{\dot{a}\sqrt{a^2+b}+a\dot{a}+\dot{b}/2}{2{\cal S}_2\sqrt{a^2+b}} \, , 
\eea
\end{subequations}
where the dot denotes derivative with respect to $z$.
Next, for any $\alpha$, $\beta$ we have
\be
\frac{\alpha}{{\cal S}_1}+\frac{\beta}{{\cal S}_2} = \alpha {\cal S}_2 +\beta {\cal S}_1 \, .
\ee 
We also find 
\be
\sqrt{a^2+b}\, {\cal S}_1 = -a{\cal S}_1+\frac{1}{{\cal S}_1} \, , \qquad \sqrt{a^2+b}\, {\cal S}_2 = a{\cal S}_2+\frac{b}{{\cal S}_2} \, . 
\ee
Consequently, combining these relations yields
\be
\sqrt{a^2+b}\,(\alpha {\cal S}_1 +\beta {\cal S}_2) = (-a\alpha+b\beta){\cal S}_1+(\alpha+a\beta){\cal S}_2 \, .
\ee
The significance of this relation is that the square root is "absorbed" into ${\cal S}_1$ and ${\cal S}_2$, such that only polynomials appear as prefactors of ${\cal S}_1$ and ${\cal S}_2$. 
Putting all this together yields
\bea
H_1\dot{\cal S}_1+H_2\dot{\cal S}_2 &=& \frac{1}{2b(a^2+b)}\left\{\left[H_1b\left(a\dot{a}-\frac{\dot{b}}{2}\right)-H_1a^2\dot{b}+H_2b\left(\dot{a}b-\frac{a\dot{b}}{2}\right)\right]{\cal S}_1 \right. \non[2ex]
&&\left. \hspace{2cm} +\left[H_2b\left(a\dot{a}+\frac{\dot{b}}{2}\right)-H_1\left(\dot{a}b-\frac{a\dot{b}}{2}\right)\right]{\cal S}_2\right\} \, .
\eea
Therefore, for the function
\be 
Q_{\rm int} = \frac{H_1 {\cal S}_1+H_2{\cal S}_2}{(a^2+b)^{n/2}b^m} \, , 
\ee
with polynomials $H_1$, $H_2$, we can write the derivative as
\bea
\dot{Q}_{\rm int} &=& \frac{h_1 {\cal S}_1+h_2{\cal S}_2}{(a^2+b)^{n/2+1}b^{m+1}} \, , 
\eea
with the polynomials
\begin{subequations} \label{polyq1q2}
\bea
h_1 &=& -H_1\left[\left(n-\frac{1}{2}\right)\left(a\dot{a}+\frac{\dot{b}}{2}\right)b+\left(m+\frac{1}{2}\right)(a^2+b)\dot{b}\right] \non[2ex]
&& +\dot{H}_1 b(a^2+b) + \frac{H_2 b}{2}\left(\dot{a}b-\frac{a\dot{b}}{2}\right) \, , \\[2ex]
h_2 &=& -H_2\left[\left(n-\frac{1}{2}\right)\left(a\dot{a}+\frac{\dot{b}}{2}\right)b+m(a^2+b)\dot{b}\right] \non[2ex]
&&+\dot{H}_2 b(a^2+b) - \frac{H_1}{2}\left(\dot{a}b-\frac{a\dot{b}}{2}\right) \, .
\eea
\end{subequations}
Now, by identifying $\dot{Q}_{\rm int}$ with $q_{\rm int}$, with $n=3$ and $m=1$, we find an analytic expression for the integral over $q_{\rm int}$: the left-hand sides of Eqs.\ (\ref{polyq1q2}) are given by the polynomials from Eqs.\ (\ref{p1p2}). Then, making a polynomial ansatz for $H_1$, $H_2$, we have reduced the problem to a system of linear equations for the coefficients of these polynomials,
which can easily be solved. The result is Eq.\ (\ref{Psi}) in the main text. 
From this result one can straightforwardly check that $\dot{Q}_{\rm int} = 2q_{\rm int}$ (the factor 2 appears due to the requirement that $q(u)$ be normalized to one in the domain $u\in [u_c,\infty]$).

\section{Derivatives of the free energy}
\label{app:mini}

Here we explain the derivation of Eqs.\ (\ref{minimOm}) in some detail since in particular taking the derivative with
respect to $u_c$ is somewhat complicated. 
We start by rewriting the free energy (\ref{Omega}) as 
\bea \label{Omeg1}
\Omega  &=& \int_{u_c}^\infty du\,\left[u^{5/2}\sqrt{(1+u^3f_Tx_4'^2-\hat{a}_0'^2+g_1)(1+g_2)} +n_I\hat{a}_0'Q\right] -\mu n_I \non[2ex]
&=& \int_{u_c}^\infty du\,u^{5/2} \eta(u) +\frac{\ell}{2}k-\mu n_I \,,  
\eea
where we have employed partial integration with the boundary values $\hat{a}_0(\infty)=\mu$, $x_4(\infty)-x_4(u_c)=\ell/2$,  have inserted the solutions of the equation of motion (\ref{a0x4}), and have 
abbreviated 
\be \label{eta}
\eta(u) \equiv \sqrt{1+g_1}\sqrt{1+g_2-\frac{k^2}{u^8 f_T}+\frac{(n_IQ)^2}{u^5}} \, .
\ee
Let us consider the free energy as a function 
\be \label{Om0}
\Omega = \Omega(u_c,k,n_I,\rho,\gamma,d) \, , 
\ee
where $\rho$, $\gamma$ are functions of $u_c$, see Eq.\ (\ref{rhogam}), 
\be
\rho = {\rm const.}\times u_c^{3/4} \, ,\qquad \gamma = {\rm const.}\times u_c^{3/2} \, .
\ee
and where $d$ is a function of $\rho, \gamma, n_I$, see Eq.\ (\ref{d}), 
\be
d = {\rm const.}\times\frac{\gamma}{\rho n_I^{1/3}} \, .
\ee
The stationary points with respect to $k$, $n_I$, and $u_c$ then are
\begin{subequations}\label{mini}
\bea 
0 &=& \frac{\partial \Omega}{\partial k} \\[2ex]
0&=&  \frac{\partial \Omega}{\partial n_I} -\frac{d}{3n_I} \frac{\partial \Omega}{\partial d} \\[2ex]
0&=&  \frac{\partial \Omega}{\partial u_c} +\frac{3\rho}{4u_c} \frac{\partial \Omega}{\partial \rho} +\frac{3\gamma}{2u_c} \frac{\partial \Omega}{\partial \gamma} + \frac{3d}{4u_c}\frac{\partial \Omega}{\partial d} \, , \label{dOmduc0}
\eea
\end{subequations} 
where the partial derivatives are taken with all other variables from Eq.\ (\ref{Om0}) held fixed. The first two equations can be straightforwardly evaluated. By using the free energy in the form of the 
second line of Eq.\ (\ref{Omeg1}) we find
\begin{subequations}
\bea 
0 &=&  \int_{u_c}^\infty du\, u^{5/2} \frac{\partial \eta}{\partial k} + \frac{\ell}{2} \, , \label{ml} \\[2ex]
0&=&   \int_{u_c}^\infty du\, u^{5/2}\left(\frac{\partial \eta}{\partial n_I} -\frac{d}{3n_I} \frac{\partial \eta}{\partial d}\right) - \mu \, , \label{mmu}
\eea
\end{subequations} 
where
\begin{subequations}
\bea
\frac{\partial \eta}{\partial k} &=& -\frac{x_4'}{u^{5/2}} \, , \\[2ex]
\frac{\partial \eta}{\partial n_I} &=& \frac{g_1\zeta^{-1}+g_2\zeta}{2n_I}+\frac{\zeta n_I Q^2}{u^5} \, , \\[2ex]
\frac{\partial \eta}{\partial d} &=& \frac{g_1\zeta^{-1}+g_2\zeta}{2q}\frac{\partial q}{\partial d} + \frac{\zeta n_I^2 Q}{u^5}\frac{\partial Q}{\partial d} \, .
\eea
\end{subequations}
This gives Eqs.\ (\ref{dk}) and (\ref{dnI}) in the main text. The derivative with respect to $u_c$ in Eq.\ (\ref{dOmduc0}) can in principle be taken directly from Eq.\ (\ref{Omeg1}). This produces a term $u_c^{5/2}\eta(u\to u_c)$, which is divergent. This divergence is eventually canceled by another term, such that there is nothing wrong with this evaluation, which has in fact been used in Refs.\ \cite{Li:2015uea,Preis:2016fsp}. However, to avoid dealing with divergent terms we may evaluate the derivative with respect to $u_c$ in the following equivalent, but more elegant, way. 
We introduce rescaled quantities by $k = \bar{k} u_c^4$, $n_I = \bar{n}_I u_c^{5/2}$, $\rho=\bar{\rho}u_c$, such that 
\be\label{Omrescaled}
\Omega = \Omega(u_c,\bar{k} u_c^4,\bar{n}_I u_c^{5/2},\bar{\rho}u_c,\gamma,d) \, , 
\ee
and by the chain rule the derivative with respect to $u_c$ at fixed $\bar{k}, \bar{n}_I, \bar{\rho}, \gamma_0$ is 
\bea \label{dOm2}
\left.\frac{\partial \Omega}{\partial u_c}\right|_{\bar{k}, \bar{n}_I, \bar{\rho},\gamma_0} &=& \frac{\partial \Omega}{\partial u_c} +\frac{4k}{u_c} \frac{\partial \Omega}{\partial k}+\frac{5n_I}{2u_c} \frac{\partial \Omega}{\partial n_I}+ \frac{\rho}{u_c} \frac{\partial \Omega}{\partial \rho} +\frac{3\gamma}{2u_c} \frac{\partial \Omega}{\partial \gamma} -\frac{d}{3u_c}\frac{\partial \Omega}{\partial d} \non[2ex]
&=& \frac{1}{4u_c}\left(\rho \frac{\partial \Omega}{\partial \rho} -d\frac{\partial \Omega}{\partial d}\right) \, ,
\eea
where, to derive the second line, Eqs.\ (\ref{mini}) have been used.
The free energy from Eq.\ (\ref{Omrescaled}) reads, after changing the integration variable to 
$\bar{u}=u/u_c$ and introducing the cutoff $\Lambda$, 
\be
\Omega = u_c^{7/2}\int_{1}^{\Lambda/u_c} d\bar{u}\,\bar{u}^{5/2} \eta(u_c\bar{u}) +\frac{\ell}{2}\bar{k}u_c^4-\mu \bar{n}_I u_c^{5/2} \, .
\ee
This expression is used to derive an explicit expression for the derivative on the left-hand side of Eq.\ (\ref{dOm2}),
\bea \label{dOmuc2}
\left.\frac{\partial \Omega}{\partial u_c}\right|_{\bar{k}, \bar{n}_I, \bar{\rho},\gamma_0} &=& \frac{7}{2}u_c^{5/2}\int_{1}^{\Lambda/u_c} d\bar{u}\,\bar{u}^{5/2} \eta(\bar{u}) - \frac{\Lambda^{7/2}\eta(\Lambda)}{u_c} +2\ell \bar{k} u_c^3-\frac{5}{2}\mu \bar{n}_I u_c^{3/2} \non[2ex]
&&+u_c^{7/2}\int_{1}^{\Lambda/u_c} d\bar{u}\,\bar{u}^{5/2}\left(\frac{\partial \eta}{\partial f_T}\frac{\partial f_T}{\partial u_c}-\frac{d}{3u_c}\frac{\partial \eta}{\partial d}\right)  \non[2ex]
&=& \frac{7}{2u_c}\left(\Omega -\frac{2}{7}\Lambda^{7/2}+\frac{st}{7}+\frac{\ell k}{14}+\frac{2\mu n_I}{7} -\frac{2}{21}\int_{u_c}^\infty du\, u^{5/2}d\frac{\partial \eta}{\partial d} \right) \, . 
\eea
As a consequence of the rescaling, the only $u_c$ dependence of $\eta$ is implicit in $f_T$ and $d$. 
In the last step, we have used $\eta(\Lambda)\to 1 $ for $\Lambda\to \infty$, and the explicit derivative
(taken after rescaling) 
\be
\frac{\partial f_T}{\partial u_c} = \frac{3 u_T^3}{u_c u^3} \, , 
\ee
which then has been used to rewrite 
\be
u_c^{7/2}\int_{1}^{\Lambda/u_c} d\bar{u}\,\bar{u}^{5/2}\frac{\partial \eta}{\partial f_T}\frac{\partial f_T}{\partial u_c} =  \frac{st}{2u_c} \, ,
\ee
with the entropy density $s$ from Eq.\ (\ref{entropy}). 
Therefore, putting together Eqs.\ (\ref{dOm2}) and (\ref{dOmuc2}) shows that stationarity with respect to $u_c$ is equivalent to the equation 
\be \label{miniuc}
\frac{7}{2}\left(\Omega -\frac{2}{7}\Lambda^{7/2} +\frac{\ell k}{14}+\frac{2\mu n_I}{7}+\frac{st}{7}-\frac{2d}{21}\frac{\partial \Omega}{\partial d} \right)=  \frac{1}{4}\left(\rho \frac{\partial \Omega}{\partial \rho} -d\frac{\partial \Omega}{\partial d}\right)\, .
\ee
This result generalizes Eq.\ (2.51) of Ref.\ \cite{Preis:2016fsp} to nonzero temperatures, nonzero interaction, and 
takes into account the different $u_c$ dependence of $\rho$. Here, $\rho\propto u_c^{3/4}$, due to Eq.\ (\ref{rhogam}), as opposed to $\rho\propto u_c$ in Ref.\ \cite{Preis:2016fsp}, which was chosen for simplicity; with $\rho\propto u_c$ the right-hand side of Eq.\ (\ref{miniuc}) would have been zero.
We now insert $\Omega$ from the second line of Eq.\ (\ref{Omeg1}), $\ell$ from Eq.\ (\ref{ml}), and $\mu$ from Eq.\ (\ref{mmu}) into the left-hand side of Eq.\ (\ref{miniuc}) to arrive at Eq.\ (\ref{duc0}) in the main text.

\section{Derivation of general expression for speed of sound}
\label{app:soundgeneral}

 The expression for the speed of sound squared $\frac{\partial P}{\partial \epsilon}$ is standard and can be found in many textbooks.  The purpose of this appendix is to derive the right-hand side of Eq.\ (\ref{sound}) and show its equivalence to $\frac{\partial P}{\partial \epsilon}$. We do so by computing the sound mode of an ideal fluid from relativistic hydrodynamics. 
We start from the general forms of the current $j^\mu$ and the stress-energy tensor $T^{\mu\nu}$ 
\bea
j^\mu &=& nu^\mu \, , \qquad 
T^{\mu\nu} = (\epsilon +P) u^\mu u^\nu - g^{\mu\nu} P \, , \label{jT}
\eea
with the number density $n$, the pressure $P$, the energy density $\epsilon$, the metric 
tensor $g^{\mu\nu} = {\rm diag}(1,-1,-1,-1)$, and the four-velocity $u^\mu = \gamma (1,\vec{u})$, where $\gamma=(1-u^2)^{-1/2}$ is the usual Lorentz factor with the modulus of the 
three-velocity $u=|\vec{u}|$. The sound modes are obtained from the hydrodynamic 
conservation equations 
\be
\partial_\mu j^\mu = \partial_\mu T^{\mu\nu} = 0 \, .
\ee
Using $dP=nd\mu+sdT$ and $\epsilon+P=\mu n+ sT$, where $s$ is the entropy density, we express all derivatives in terms of 
derivatives of the independent variables four-velocity $u^\mu$, chemical potential $\mu$, and temperature $T$,
\begin{subequations}
\bea
\partial_\mu j^\mu &=& n\partial\cdot u+\frac{\partial n}{\partial \mu}(u\cdot\partial)\mu+\frac{\partial s}{\partial \mu}(u\cdot\partial) T \, , \\[2ex]
\partial_\mu T^{\mu\nu} &=& \mu u^\nu \partial_\mu j^\mu +\mu n(u\cdot\partial)u^\nu +sT[u^\nu\partial\cdot u+(u\cdot\partial)u^\nu] \\[2ex]
&&+\left[\left(n+T\frac{\partial n}{\partial T}\right) u^\nu u\cdot\partial -n\partial^\nu\right] \mu + \left[\left(s+T\frac{\partial s}{\partial T}\right) u^\nu u\cdot\partial -s\partial^\nu\right] T
 \,. 
\eea
\end{subequations}
(By contracting the second equation with $u_\nu$ one obtains the equation for the entropy current, which is conserved in the present non-dissipative case. But for our purpose this reformulation is not necessary.) We now introduce fluctuations in the three-velocity as well as in the chemical potential and temperature, $\vec{u}(\vec{x},t) = \delta\vec{u}\, e^{i(\omega t-\vec{k}\cdot\vec{x})}$, $\mu(\vec{x},t) = \mu + \delta\mu \, e^{i(\omega t-\vec{k}\cdot\vec{x})}$, $T(\vec{x},t) = T+ \delta T\,  e^{i(\omega t-\vec{k}\cdot\vec{x})}$, where $\mu$ and $T$ are equilibrium values (we do not impose an external velocity $\vec{u}$ on the fluid). Linearizing  in the fluctuations yields
\begin{subequations}
\bea
\partial_\mu j^\mu &\simeq & ie^{i(\omega t-\vec{k}\cdot\vec{x})}\left[\omega\left(\frac{\partial n}{\partial \mu}\delta\mu + \frac{\partial s}{\partial\mu}\delta T\right) -n\vec{k}\cdot\delta\vec{u}\right]=0 \, , \\[2ex]
\partial_\mu T^{\mu 0} &\simeq&  
ie^{i(\omega t-\vec{k}\cdot\vec{x})}\left[\omega T\left(\frac{\partial n}{\partial T}\delta\mu +\frac{\partial s}{\partial T}\delta T\right)-sT\vec{k}\cdot\delta\vec{u}\right]
+\mu\partial_\mu j^\mu =0 \, , \\[2ex]
\partial_\mu T^{\mu i} &\simeq& -ie^{i(\omega t-\vec{k}\cdot\vec{x})}\Big[k_i(n\delta\mu+s\delta T)-w\omega\delta u_i \Big] =0\, ,
\eea
\end{subequations}
where $w= \mu n+sT$ is the enthalpy density. 
We contract the last equation with $k_i$ to arrive at three equations for the 
three fluctuations $\delta \mu$, $\delta T$, $\vec{k}\cdot\delta\vec{u}$. (Contracting with 
$\delta_{ij} -\hat{k}_i\hat{k}_j$ yields a decoupled equation resulting in the trivial transverse mode $\omega=0$.)  
Nontrivial solutions are found by setting the determinant of the corresponding $3\times 3$ matrix to zero. This 
determinant is a polynomial in $\omega$ of degree 3 and we find the solutions $\omega=0$ and 
\be
\omega = \pm c_s k 
\ee
with 
\be
c_s^2 = \frac{n^2\frac{\partial s}{\partial T}+s^2\frac{\partial n}{\partial\mu}-ns\left(\frac{\partial n}{\partial T}+\frac{\partial s}{\partial \mu}\right)}{w\left(\frac{\partial n}{\partial \mu}\frac{\partial s}{\partial T}-  
\frac{\partial n}{\partial T}\frac{\partial s}{\partial \mu}\right)} \, . \label{cs3}
\ee
This is the right-hand side of Eq.\ (\ref{sound}).

We can rewrite this result by first changing variables 
from ($\mu,T$) to ($n,s$). The partial derivatives are 
related by the Jacobian of the map $f(\mu,T)=[n(\mu,T),s(\mu,T)]$, such that we can simplify Eq.\ (\ref{cs3}) to 
\be
c_s^2 = \frac{1}{w}\left[n\left(n\frac{\partial\mu}{\partial n}+s \frac{\partial T}{\partial n}\right)+s\left(n\frac{\partial\mu}{\partial s}+s \frac{\partial T}{\partial s}\right)\right]
=\frac{1}{w}\left(n\frac{\partial P}{\partial n} + s\frac{\partial P}{\partial s}\right) \, , \label{cs4}
\ee
where, in the second step, we have used  $P=P[\mu(n,s),T(n,s)]$. All partial derivatives with respect to $n$ are taken at fixed $s$ and vice versa. Next, we change variables from $(n,s)$ to $(\epsilon,s/n)$, where $s/n$ is the entropy per particle. We use $d\epsilon=Tds+\mu dn$ and
$d(s/n)=ds/n-sdn/n^2$ to find 
\be
dn=\frac{n}{w}\left[d\epsilon -nTd(s/n)\right] \, , \qquad ds = \frac{1}{w}[sd\epsilon + \mu n^2d(s/n)] \, ,
\ee
and thus 
\bea
dP =  \frac{1}{w}\left(n\frac{\partial P}{\partial n} + s\frac{\partial P}{\partial s}\right) d\epsilon+
 \frac{n^2}{w}\left(-T\frac{\partial P}{\partial n} + \mu\frac{\partial P}{\partial s}\right) d(s/n) \, .
\eea
As we see from Eq.\ (\ref{cs4}) the squared speed of sound 
is the coefficient in front of $d\epsilon$, and thus 
\be
c_s^2 = \frac{\partial P}{\partial \epsilon} \, ,
\ee
where the derivative is taken at fixed entropy per particle $s/n$.

\section{Speed of sound in mesonic and chirally symmetric phases}
\label{app:sound}

This appendix gives a brief derivation of the speed of sound in the mesonic phase and the chirally symmetric phase. The 
relevant free energies are taken from the literature, most conveniently from appendix B of Ref.\ \cite{Li:2015uea}, which uses the same notation as the present paper. 

\subsection{Mesonic phase}

In the mesonic phase, the location of the tip of the connected flavor branes $u_c$ is determined from 
\be
\frac{\ell}{2} = \int_{u_c}^\infty du\, x_4' \, , \qquad x_4' = \frac{u_c^4\sqrt{f_T(u_c)}}{u^{3/2}\sqrt{f_T(u)}\sqrt{u^8 f_T(u)-u_c^8f_T(u_c)}} \, .
\ee 
In general, this has to be solved numerically. For small temperatures $t$ we find 
\be \label{u0exp}
\ell^2 u_c = u_c^{(0)}+ u_c^{(1)} (\ell t)^6 \, , 
\ee
with 
\bea
u_c^{(0)} &=& 16 \pi \left[\frac{\Gamma\left(\frac{9}{16}\right)}{\Gamma\left(\frac{1}{16}\right)}\right]^2  \, , \qquad 
u_c^{(1)}  = \frac{8\pi^{7/2}}{729}   \left[\frac{\Gamma\left(\frac{1}{16}\right)}{\Gamma\left(\frac{9}{16}\right)}\right]^5 {\cal I}_1 \, , 
\eea
where we have abbreviated
\be
{\cal I}_1 \equiv \int_1^\infty du\,\frac{1+u+u^2+u^3+u^4+u^5+u^6+u^7-u^8-u^9-u^{10}}{(1+u+u^2+u^3+u^4+u^5+u^6+u^7)u^{9/2}\sqrt{u^8-1}} \simeq 0.0778 \, .
\ee
The dimensionless free energy, including the vacuum subtraction, is 
\bea
\Omega_\cup &=& \int_{u_c}^\infty du\, u^{5/2}\left[\frac{u^4\sqrt{f_T(u)}}{\sqrt{u^8f_T(u)-u_c^8f_T(u_c)}} -1\right] -\frac{2}{7}u_c^{7/2} \, .
\eea
Again, this can be evaluated semi-analytically for small temperatures. Using the expansion for $u_c$ (\ref{u0exp})
we find 
\be
\ell^7 \Omega_\cup \simeq \Omega^{(0)}+ \Omega^{(1)}(\ell t)^6 \, ,
\ee
with
\begin{subequations}
\bea
\Omega^{(0)} &=& -\frac{2^{15}\pi^4}{7}\frac{\Gamma\left(\frac{15}{16}\right)\tan\frac{\pi}{16}}{\Gamma\left(\frac{7}{16}\right)}\left[\frac{\Gamma\left(\frac{9}{16}\right)}{\Gamma\left(\frac{1}{16}\right)}\right]^7 \, , \\[2ex]
\Omega^{(1)} &=& -\frac{(2\sqrt{\pi})^{13}}{729}\left[{\cal I}_1\frac{\Gamma\left(\frac{15}{16}\right) \tan\frac{\pi}{16}}{\Gamma\left(\frac{7}{16}\right)}+{\cal I}_2 \frac{\Gamma\left(\frac{9}{16}\right)}{\Gamma\left(\frac{1}{16}\right)}\right] \, , 
\eea
\end{subequations}
where we have abbreviated 
\be
{\cal I}_2 \equiv \int_1^\infty du\,\frac{u^{7/2}(1+u+u^2)}{(1+u+u^2+u^3+u^4+u^5+u^6+u^7)\sqrt{u^8-1}} \simeq 0.18236 \, .
\ee
The zero-temperature result gives the pressure (\ref{OmM}). 
In the mesonic phase, the baryon density is zero for all temperatures, and the pressure does not depend on the chemical potential. Therefore, with Eq.\ (\ref{sound}) the speed of sound becomes
\be
c_s^2 = \frac{s}{T}\left(\frac{\partial s}{\partial T}\right)^{-1} \, . 
\ee
The above expansion shows that for small temperatures $s\propto T^5$, and thus we obtain $c_s^2(\mu,T\to 0)=1/5$, 
which is independent of the 
coefficients of the low-temperature expansion.

\subsection{Chirally symmetric phase}

In the chirally symmetric phase, again using appendix B of Ref.\ \cite{Li:2015uea}, the solution of the equation of motion for 
the temporal component of the abelian gauge field is 
\be
\hat{a}_0(u) = \mu-\frac{n_I^{2/5}\Gamma\left(\frac{3}{10}\right)\Gamma\left(\frac{6}{5}\right)}{\sqrt{\pi}}
+ u\,{}_2 F_1\left[\frac{1}{5},\frac{1}{2},\frac{6}{5},-\frac{u^5}{n_I^2}\right] \, , 
\ee
with $n_I$ determined as a function of $\mu$ and $T$ from 
\be \label{ceq}
0 = \mu-\frac{n_I^{2/5}\Gamma\left(\frac{3}{10}\right)\Gamma\left(\frac{6}{5}\right)}{\sqrt{\pi}}
+ u_T\,{}_2 F_1\left[\frac{1}{5},\frac{1}{2},\frac{6}{5},-\frac{u_T^5}{n_I^2}\right] \, .
\ee
The dimensionless free energy is
\be
\Omega_{||} = \int_{u_T}^\infty du\,\frac{u^5}{\sqrt{u^5+n_I^2}}   = \frac{2}{7}\Lambda^{7/2}-\frac{2\Gamma\left(\frac{3}{10}
\right)\Gamma\left(\frac{6}{5}\right)}{7\sqrt{\pi}}\,n_I^{7/5}-\frac{2u_T n_I}{7}h\left(\frac{u_T^{5/2}}{n_I}\right) \, ,
\ee
where
\be
h(x)\equiv \sqrt{1+x^2}-{}_2F_1\left[\frac{1}{5},\frac{1}{2},\frac{6}{5},-x^2\right] \, .
\ee
Even though $n_I$ itself does not have a simple analytic form, we can compute its derivatives and thus the derivatives of the free energy analytically 
and insert the result into the expression for the sound speed (\ref{sound}). This yields Eq.\ (\ref{cs2}) in the main text.

\bibliographystyle{JHEP}
\bibliography{references}

\end{document}